\documentclass[12pt,preprint]{emulateapj}

\newcommand{\be}{\begin{equation}}
\newcommand{\ee}{\end{equation}}
\newcommand{\ba}{\begin{eqnarray}}
\newcommand{\ea}{\end{eqnarray}}

\begin{document}
\title{Cosmology from the Chinese Space Station Optical Survey (CSS-OS)}

\author{Yan Gong$^{1*}$, Xiangkun Liu$^{2}$, Ye Cao$^{1,3}$, Xuelei Chen$^{1,3,4}$, Zuhui Fan$^{2,5}$, Ran Li$^{1,3}$, \\Xiao-Dong Li$^{6}$, Zhigang, Li$^{7,8}$, Xin Zhang$^{9}$, Hu Zhan$^{9,10}$}

\affil{$^{1}$ Key Laboratory for Computational Astrophysics, National Astronomical Observatories, \\Chinese Academy of Sciences, 20A Datun Road, Beijing 100012, China }
\affil{$^{2}$ South-Western Institute for Astronomy Research, Yunnan University, Kunming 650500, China}
\affil{$^{3}$ School of Astronomy and Space Sciences, University of Chinese Academy of Sciences, Beijing 100049, China}
\affil{$^{4}$ Centre for High Energy Physics, Peking University, Beijing 100871, China}
\affil{$^{5}$ Department of Astronomy, School of Physics, Peking University, Beijing 100871, China}
\affil{$^{6}$ School of Physics and Astronomy, Sun Yat-Sen University, Guangzhou 510297, P. R. China}
\affil{$^{7}$  Center for Astronomy and Astrophysics, Department of Physics and Astronomy, Shanghai Jiao Tong University,\\ 800 Dongchuan Road, Shanghai, 200240}
\affil{$^{8}$  IFSA Collaborative Innovation Center, Shanghai Jiao Tong University, Shanghai 200240, China}
\affil{$^{9}$ Key Laboratory of Space Astronomy and Technology, National Astronomical Observatories,\\ Chinese Academy of Sciences, Beijing 100012, China}
\affil{$^{10}$ Kavli Institute for Astronomy and Astrophysics, Peking University, Beijing 100871, China}
\email{E-mail: gongyan@bao.ac.cn}

\begin{abstract}
The Chinese Space Station Optical Survey (CSS-OS) is a planned full sky survey operated by the Chinese Space Station Telescope (CSST). It can simultaneously perform the photometric imaging and spectroscopic slitless surveys, and will probe weak and strong gravitational lensing, galaxy clustering, individual galaxies and galaxy clusters, active galactic nucleus (AGNs), and so on. It aims to explore the properties of dark matter and dark energy and other important cosmological problems. In this work, we focus on two main CSS-OS scientific goals, i.e. the weak gravitational lensing (WL) and galaxy clustering surveys. We generate the mock CSS-OS data based on the observational COSMOS and zCOSMOS catalogs. We investigate the constraints on the cosmological parameters from the CSS-OS using the Markov Chain Monte Carlo (MCMC) method. The intrinsic alignments, galaxy bias, velocity dispersion, and systematics from instrumental effects in the CSST WL and galaxy clustering surveys are also included, and their impacts on the constraint results are discussed. We find that the CSS-OS can improve the constraints on the cosmological parameters by a factor of a few (even one order of magnitude in the optimistic case), compared to the current WL and galaxy clustering surveys. The constraints can be further enhanced when performing joint analysis with the WL, galaxy clustering, and galaxy-galaxy lensing data. Therefore, the CSS-OS is expected to be a powerful survey for exploring the Universe. Since some assumptions may be still optimistic and simple, it is possible that the results from the real survey could be worse. We will study these issues in details with the help of simulations in the future.

\end{abstract}


\maketitle

\section{Introduction}

Understanding the nature of dark matter and dark energy, and formation and evolution of the cosmic large-scale structure (LSS) is essential for the study of cosmology. A number of powerful observational tools, such as weak gravitational lensing (WL) \citep[e.g.][]{Kaiser92,Kaiser98}, baryon acoustic oscillations (BAO) \citep[e.g.][]{Eisenstein05a,Eisenstein05b}, and redshift-space distortion (RSD) \citep[e.g.][]{Jackson72,Kaiser87}, have been applied for solving these issues. A few Stage IV ground- and space-borne telescopes, e.g. the Large Synoptic Survey Telescope (LSST)\footnote{\tt https://www.lsst.org/} \citep{Ivezic08,Abell09}, $Euclid$ space telescope\footnote{\tt https://www.euclid-ec.org/} \citep{Laureijs11}, and Wide Field Infrared Survey Telescope (WFIRST)\footnote{\tt https://wfirst.gsfc.nasa.gov/}, have been planned to perform these measurements. These powerful surveys are expected to make great improvements on related scientific objectives. The Chinese Space Station Optical Survey (CSS-OS) is another this kind of sky survey. 

\begin{figure}[t]
\includegraphics[scale = 0.42]{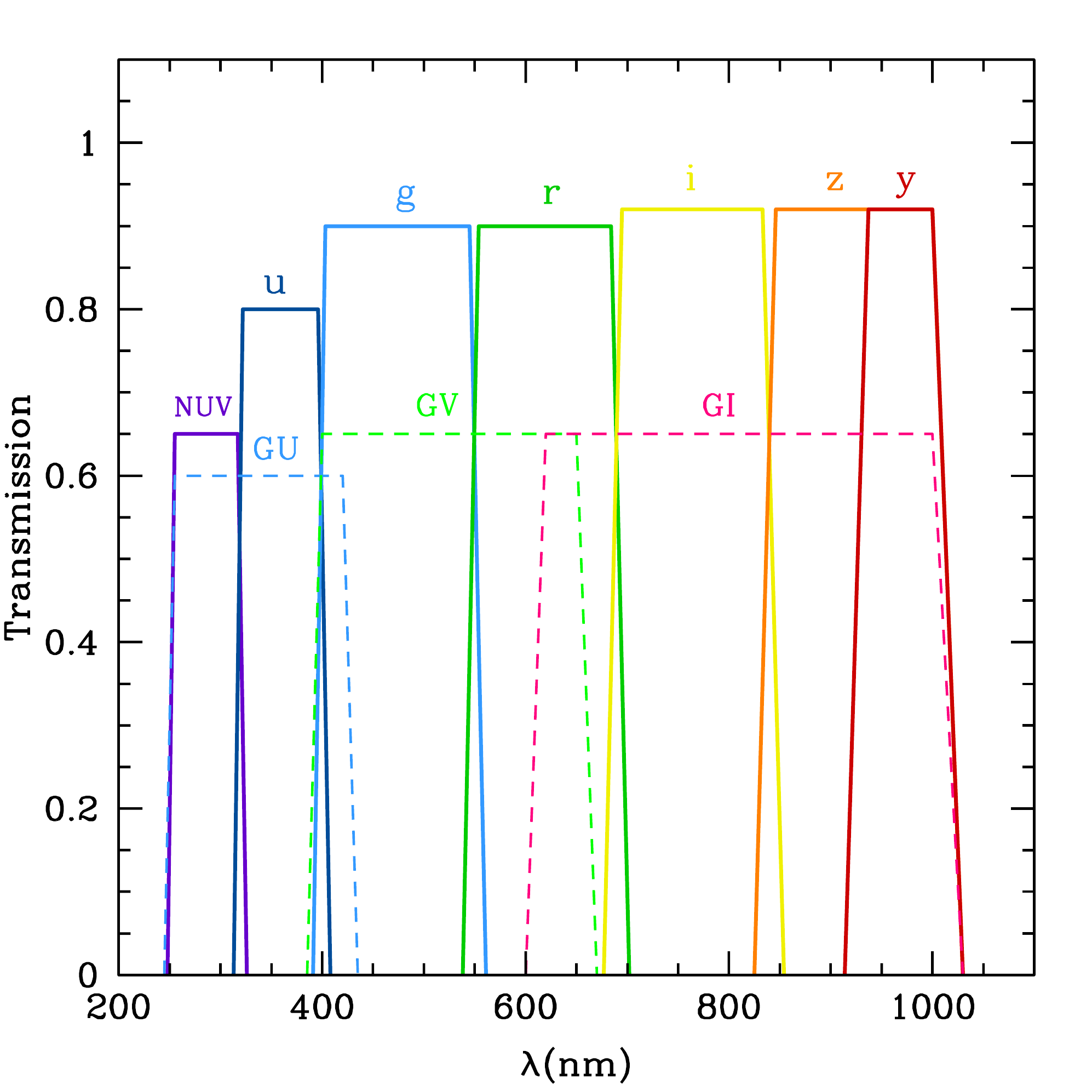}
\caption{\label{fig:filters} The intrinsic transmissions for the filters of the CSST photometric (solid) and spectroscopic (dashed) surveys. There are seven bands, i.e. $NUV$, $u$, $g$, $r$, $i$, $z$, and $y$, in the photometric imaging survey, and three wide bands, i.e. $GU$, $GV$, and $GI$, in the slitless spectroscopic survey.}
\end{figure}

The CSS-OS is a major science project established by the space application system of the China Manned Space Program, which will be performed by the Chinese Space Station Telescope (CSST) \citep{Zhan11,Zhan18,Cao18}. The CSST is a 2-meter space telescope in the same orbit of the China Manned Space Station, and is planned to be launched at the end of 2022. The CSS-OS will cover 17500 deg$^2$ sky area  in about ten years with field of view $1.1$ deg$^2$. It will simultaneously perform both photometric imaging and slitless grating spectroscopic surveys with high spatial resolution $\sim0.15''$ (80\% energy concentration region) and wide wavelength coverage. There are seven photometric imaging bands and three spectroscopic bands covering 255-1000 nm (see Table~\ref{tab:Des_paramts} and Figure~\ref{fig:filters}). A few important cosmological and astronomical objectives will be explored by the CSS-OS, such as the properties of dark matter and dark energy, cosmic large-scale structure, galaxy formation and evolution, galaxy clusters, active galactic nucleus (AGNs), etc. Comparing to other next generation surveys, the CSS-OS has several advantages, such as wider wavelength coverage, larger number of filters, smaller spatial resolution, better image quality, simultaneous photo+spec performance. Therefore, the CSS-OS is expected to be a powerful survey for probing the Universe, which is comparable and even more robust in some aspects than other Stage IV surveys \citep{Zhan11}.

In this work, we predict the measurements of weak lensing and galaxy clustering for the CSS-OS, and investigate the constraint accuracy of the cosmological parameters. We make use of two catalogs from the real observations, i.e. COSMOS and zCOSMOS surveys, as the mock catalogs for the CSST photometric and spectroscopic surveys, respectively. These two catalogs have similar magnitude limits as the CSS-OS, and can well represent the CSST observations. For the weak lensing survey, we derive the galaxy redshift distribution from the mock catalog, and divide it into several photometric redshift (photo-$z$) bins to calculate the auto and cross convergence power spectra. The galaxy intrinsic alignments and systematics (multiplicative and additive) due to point spread function, photometry offsets, instrumental noise, etc., are included when estimating the errors. We also evaluate the redshift-space galaxy clustering power spectra in the CSST spectroscopic survey with slitless gratings. The multipole power spectra are calculated in the spectroscopic redshift (spec-$z$) bins, and the effects of frequency resolution, galaxy bias, velocity dispersion, and systematic errors are considered in the error estimate. Then we compute the observed CSS-OS angular cross power spectra of the weak lensing and galaxy clustering, i.e. galaxy-galaxy lensing power spectra.  The mock data of weak lensing, galaxy clustering, and their cross correlation are used in the constraints on the cosmological parameters. The Markov Chain Monte Carlo (MCMC) technique is adopted to illustrate the probability distributions of the parameters.

The paper is organized as follows: the weak lensing, galaxy clustering, and galaxy-galaxy lensing power spectra surveys are discussed in Section 2, 3, and 4, respectively. In section 5, we show the details of fitting process using the MCMC method. The constraint results are shown in Section 6. We finally summarize the conclusions in Section 7. Throughout the paper, we assume the flat $\Lambda$CDM cosmology with $\Omega_{\rm m}=0.3$, $\Omega_{\rm b}=0.05$, $\sigma_8=0.8$, $n_{\rm s}=0.96$, and $h=0.7$ as the fiducial model.

\begin{table*}
\begin{center}
\caption{Designed and assumed parameters of the CSST photometric and spectroscopic surveys.}
\label{tab:Des_paramts}
\vspace{1mm}
\begin{tabular}{ l | l | l | l | l | l | l | | | | | | | | | l | l | l | l | l | l | l | l | l | l | l | l | l | | |}
\hline\hline
\multicolumn{8}{c} {Survey Characteristics} \vspace{1mm} \\
\hline
\multicolumn{1}{c|} {Telescope} & \multicolumn{7}{c} {2 m primary, off-axis TMA, orbit $\sim$ 400 km, duration $\sim$ 10 yrs} \\
\hline
\multicolumn{1}{c|} {Survey mode} & \multicolumn{7}{c} {Photometric imaging + slitless spectroscopic joint survey} \\
\hline
\multicolumn{1}{c|} {Wide survey} &  \multicolumn{7}{c} {15000 deg$^2$ ($|b|\gtrsim$20$^{\circ}$) + 2500  deg$^2$ ($15^{\circ}\lesssim|b|<$20$^{\circ}$)}\\
\hline
\multicolumn{1}{c|} {Deep survey} &  \multicolumn{7}{c} {400 deg$^2$ (selected areas)}\\
\hline
\multicolumn{1}{c|} {Field of view} &  \multicolumn{7}{c} {$\gtrsim$ 1.1 deg$^2$}\\
\hline
\multicolumn{8}{c} {Photometric wide survey} \vspace{1mm}  \\
\hline
\multicolumn{1}{c|} {Bands} & \multicolumn{1}{|c|}{$NUV$} & \multicolumn{1}{|c|}{$u$} &\multicolumn{1}{|c|}{$g$} & \multicolumn{1}{|c|}{$r$} & \multicolumn{1}{|c|}{$i$} & \multicolumn{1}{|c|}{$z^*$} &\multicolumn{1}{|c}{$y^*$}  \\
\hline
\multicolumn{1}{c|} {Wavelength ($\lambda_{\rm -90}-\lambda_{\rm +90}$ nm)} & \multicolumn{1}{|c|}{$255-317$} & \multicolumn{1}{|c|}{$322-396$} &\multicolumn{1}{|c|}{$403-545$} & \multicolumn{1}{|c|}{$554-684$} & \multicolumn{1}{|c|}{$695-833$} & \multicolumn{1}{|c|}{$846-1000$} &\multicolumn{1}{|c}{$937-1000$}  \\
\hline
\multicolumn{1}{c|} {Exposure time} & \multicolumn{1}{|c|}{$150$\,s $\times4$} & \multicolumn{1}{|c|}{$150$\,s $\times2$} &\multicolumn{1}{|c|}{$150$\,s $\times2$} & \multicolumn{1}{|c|}{$150$\,s $\times2$} & \multicolumn{1}{|c|}{$150$\,s $\times2$} & \multicolumn{1}{|c|}{$150$\,s $\times2$} &\multicolumn{1}{|c}{$150$\,s $\times4$}  \\
\hline
\multicolumn{1}{c|} {Sensitivity (point/extended$^a$)} & \multicolumn{1}{|c|}{$25.4/24.2$} & \multicolumn{1}{|c|}{$25.4/24.2$} &\multicolumn{1}{|c|}{$26.3/25.1$} & \multicolumn{1}{|c|}{$26.0/24.8$} & \multicolumn{1}{|c|}{$25.9/24.6$} & \multicolumn{1}{|c|}{$25.2/24.1$} &\multicolumn{1}{|c}{$24.4/23.2$}  \\
\hline
\multicolumn{1}{c|} {PSF size ($\le R_{\rm EE80}$/FWHM$^b$)} & \multicolumn{1}{|c|}{$0.15''/0.20''$} & \multicolumn{1}{|c|}{$0.15''/0.20''$} &\multicolumn{1}{|c|}{$0.15''/0.20''$} & \multicolumn{1}{|c|}{$0.15''/0.20''$} & \multicolumn{1}{|c|}{$0.16''/0.21''$} & \multicolumn{1}{|c|}{$0.18''/0.24''$} &\multicolumn{1}{|c}{$0.18''/0.24''$}  \\
\hline
\multicolumn{8}{c} {Spectroscopic wide survey (gratings)} \vspace{1mm}  \\
\hline
\multicolumn{1}{c|} {Bands} & \multicolumn{2}{|c|}{$GU$} & \multicolumn{2}{|c|}{$GV$} &\multicolumn{3}{|c}{$GI^*$}\\
\hline
\multicolumn{1}{c|} {Wavelength ($\lambda_{\rm -90}-\lambda_{\rm +90}$ nm)} & \multicolumn{2}{|c|}{$255-420$} & \multicolumn{2}{|c|}{$400-650$} &\multicolumn{3}{|c}{$620-1000$}\\
\hline
\multicolumn{1}{c|} {Exposure time} & \multicolumn{2}{|c|}{$150$\,s $\times4$} & \multicolumn{2}{|c|}{$150$\,s $\times4$} &\multicolumn{3}{|c}{$150$\,s $\times4$}\\
\hline
\multicolumn{1}{c|} {Sensitivity$^c$ [mag / (erg$/$s$/$cm$^2)$]} & \multicolumn{2}{|c|}{$20.5$ / $1.0\times10^{-15}$} & \multicolumn{2}{|c|}{$21.0$ / $4.0\times10^{-16}$} &\multicolumn{3}{|c}{$21.0$ / $2.6\times10^{-16}$}\\
\hline
\multicolumn{1}{c|} {Spectral resolution $R$ $^d$} & \multicolumn{2}{|c|}{$\ge 200$} & \multicolumn{2}{|c|}{$\ge 200$} &\multicolumn{3}{|c}{$\ge 200$}\\
\hline
 \end{tabular}
\end{center}
\vspace{-2mm}
$^*$ cut off by detector quantum efficiency. \\
$^a$ 5$\sigma$ AB mag, and assuming galaxies with 0.3$''$ half-light radius for extended sources. In the CSST deep survey, the sensitivity can be 1 mag deeper at least than the wide survey. The corresponding SNR cut is $\sim 5$ for the $i$ band. The magnitude limits shown here are the updates of the values given in \cite{Cao18}.\\
$^b$ Assuming Gaussian-like PSF profile, and $R_{\rm EE80}$ is the radius of 80\% energy concentration. This PSF size indicates that the peak of galaxy size distribution is $\sim 0.2''$, and it has a cut around $\sim 0.13''$.\\
$^c$ 5$\sigma$ AB mag per resolution element for point sources. In the CSST deep survey, the sensitivity can be 1 mag deeper  at least than the wide survey. Note that since the region of emitting lines in an emission line galaxy (ELG) is usually small compared to the full size of galaxy, we treat the ELGs as point sources here for simplicity.\\
$^d$ $R=\lambda/{\Delta \lambda}$.
\end{table*}


\section{Weak lensing survey}

The CSS-OS covers 17500 deg$^2$ survey area in the photometric imaging survey with survey depth $i\simeq26$ AB magnitude (5$\sigma$ detection for point sources)\citep{Cao18}. It has seven filters, i.e. $NUV$, $u$, $g$, $r$, $i$, $z$, and $y$ bands, which covers the wavelength range 255-1000 nm (see Figure~\ref{fig:filters}). The point spread function (PSF) of the CSS-OS is designed to be as small as 0.15'' within 80\% energy concentration, and it requires the maximum PSF ellipticity is less than 0.15 at any position of the field of view (FoV), and the average value is less than 0.05. Hence it is expected to obtain excellent galaxy shape measurements for the weak lensing study. In this section, we discuss the predicated galaxy redshift distribution and shear power spectra for the CSS-OS.

\subsection{Galaxy photometric redshift distribution}

\begin{figure}[t]
\includegraphics[scale = 0.42]{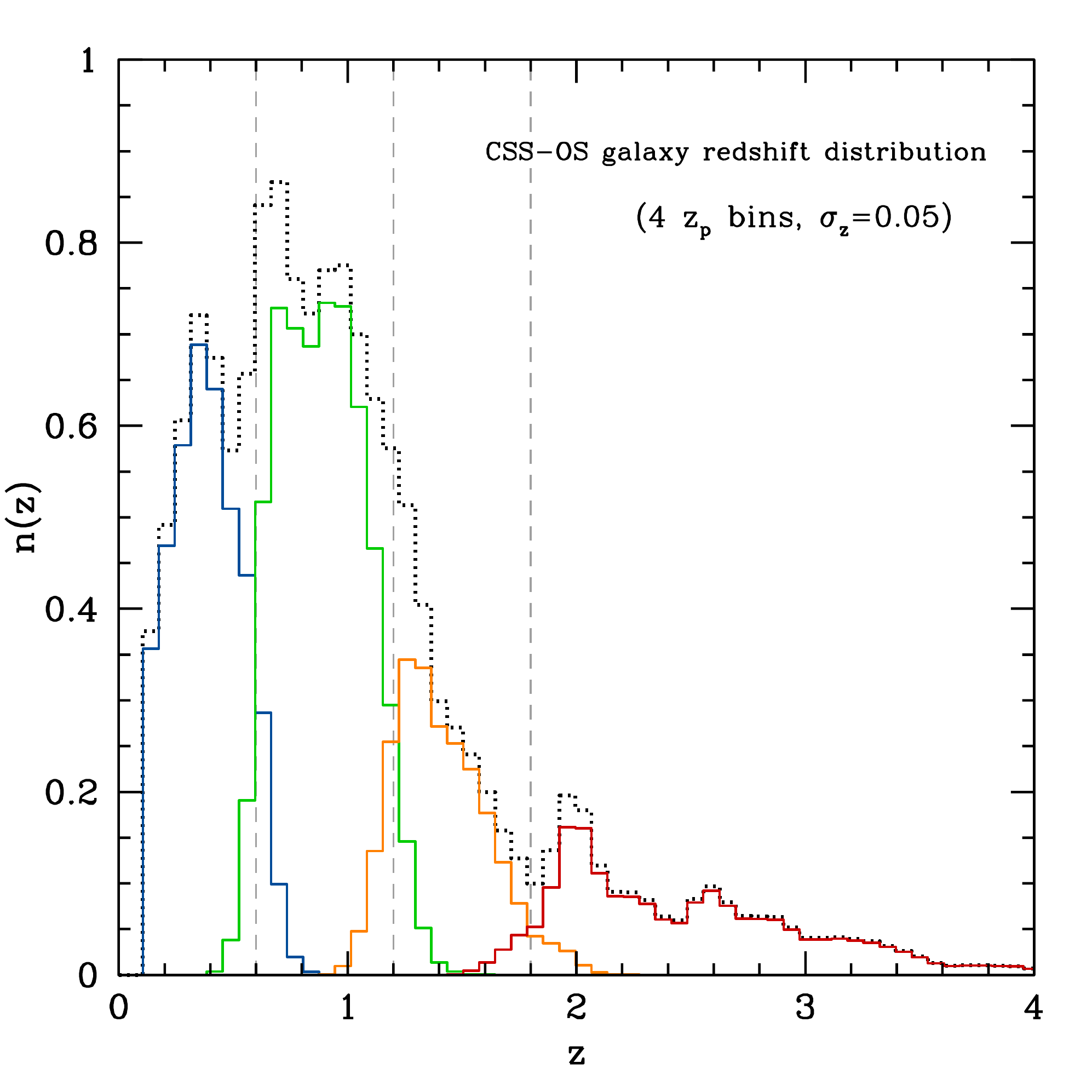}
\caption{\label{fig:nz_phot} The mock galaxy redshift distributions in the photometric imaging survey of the CSS-OS. The black dotted line denotes the total redshift distribution $n(z)$, which is obtained from the COSMOS catalog \citep{Cao18}. The solid blue, green, orange, and red lines are the redshift distribution $n_i(z)$ for the four photo-$z$ bins (divided by the gray vertical dashed lines) with ${\Delta z}=0$ and $\sigma_z=0.05$ in Eq.~(\ref{eq:pz}).}
\end{figure}

Following \cite{Cao18}, we adopt a galaxy redshift distribution $n(z)$ derived from the COSMOS catalog for the CSS-OS photometric survey \citep{Capak07, Ilbert09}. This catalog contains about 220,000 galaxies in 2 deg$^2$, and has similar magnitude limit as the CSS-OS with $i\le25.2$ for galaxy observation. Although the survey area of the CSS-OS is much larger, it can represent the redshift and magnitude distributions, and galaxy types observed by the CSS-OS. The CSS-OS redshift distribution derived from this catalog is shown in dotted line in Figure~\ref{fig:nz_phot}. We can find that the redshift distribution has a peak around $z=0.6$, and can extend to $z\sim4$.

In order to extract more information from the weak lensing data, we divide the redshift range into different photo-$z$ tomographic bins, and study the auto and cross power spectra of these bins. As shown in Figure~\ref{fig:nz_phot}, for instance, we divide the redshift range into four photo-$z$ bins (gray vertical lines). The first three bins has equal interval with $\Delta_{z}=0.6$, and the last bin occupies the rest of the redshift range of redshift distribution\footnote{There are several binning strategies can be adopted in the weak lensing tomographic studies, such as equal binning, $\Delta_z\sim 1+z$ binning, etc. We find that the results would not be sensitive to the binning strategy, as long as there is not much difference (in one order of magnitude) of the average source number densities between the bins.}. Note that the fitting results of the cosmological parameters also depends on the number of photo-$z$ bins \citep[e.g.][]{Huterer02}. Under the assumptions of the systematics as discussed in Section 2.2, we find that although more number of bins may further improve the constraint results, the improvement is not much significant (averagely a factor of $\sim$1.3 and $\sim$1.5 on the standard deviations of cosmological parameters for the five and six photo-$z$ bins cases, respectively). Considering the purpose of this work, a four-bin division is adequate for our study. More number of bins can be used in the real data analysis, which may help to improve the constraints on the cosmological parameters.

The real galaxy redshift distribution in the $i$th photo-z bin can be expressed as \citep[e.g.][]{Ma06}
\be \label{eq:nzi}
n_i(z) = \int_{z_{\rm p,l}^i}^{z_{\rm p,u}^{i}} {\rm d}z_{\rm p}\ n(z)\ p(z_{\rm p}|z),
\ee
where $z_{\rm p,l}^i$ and $z_{\rm p,u}^i$ are the lower and upper limits of the $i$th photo-$z$ bin, $n(z)$ is the total redshift distribution, and $p(z_{\rm p}|z)$ is the photo-$z$ distribution function given the real redshift $z$. We assume it takes the form as
\be \label{eq:pz}
p(z_{\rm p}|z) = \frac{1}{\sqrt{2\pi}\sigma_z}\ {\rm exp}\left[ -\frac{(z-z_{\rm p}-{\Delta z})^2}{2\sigma_z^2} \right],
\ee
where ${\Delta z}$ and $\sigma_z$ are the redshift bias and scatter, respectively, which vary as functions of redshift. In this work, we assume that they are constants in different photo-$z$ bins, and treat them as free parameters when fitting the data. Then the $n_i(z)$ in Eq.~(\ref{eq:nzi}) can be reduced to
\be
n_i(z) = \frac{1}{2}n(z)\left[{\rm erf}(x_{\rm u}^i)-{\rm erf}(x_{\rm l}^i)\right],
\ee
where ${\rm erf}(x)$ is the error function, and 
\be
x^i_{\rm u/l}=\left( z_{\rm p, u/l}^i-z+{\Delta z}\right)/\sqrt{2}\sigma_z.
\ee
Note that we always have $n(z)=\sum_i n_i(z)$, no matter what form $n_i(z)$ takes. 

In Figure~\ref{fig:nz_phot}, the solid blue, green, orange, and red lines show the $n_i(z)$ for the four photo-$z$ bins with ${\Delta z}=0$ and $\sigma_z=0.05$. In the COSMOS catalog, we find that there are about 208,000 galaxies with $\sigma_z\le0.05$, which take about 95\% of the whole sample \citep{Cao18}. This number can be used, as a reference, to estimate the galaxy number density in the CSS-OS weak lensing survey. We will adopt the $n_i(z)$ shown in Figure~\ref{fig:nz_phot} and the estimated number density in the following WL discussion.

\subsection{Shear power spectra}

\begin{figure}[t]
\includegraphics[scale = 0.42]{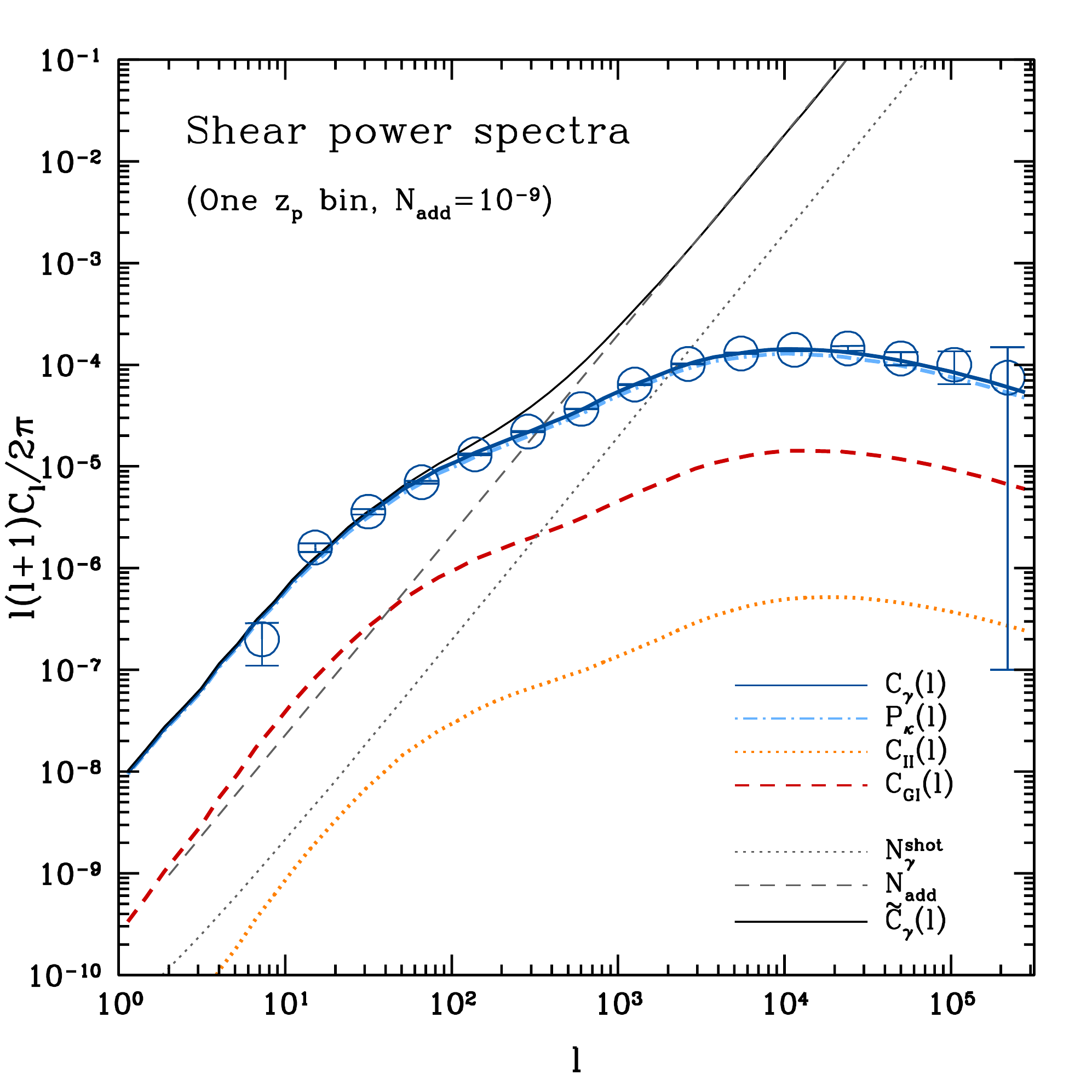}
\caption{\label{fig:P_WL} The components of the shear power spectrum measured by the CSS-OS for the one-bin case assuming $\bar{N}_{\rm add}=10^{-9}$. The blue solid, light blue dash-dotted, orange dotted, and red dashed curves denote the signal, convergence, intrinsic-intrinsic, gravitational-intrinsic power spectra, respectively. The shot-noise and additive term are shown in black dotted and dashed curves, respectively. The black solid line is for the total power spectrum. The blue circles with error bars are the mock data for the signal power spectrum.}
\end{figure}

\begin{figure*}
\centerline{
\resizebox{!}{!}{\includegraphics[scale=0.8]{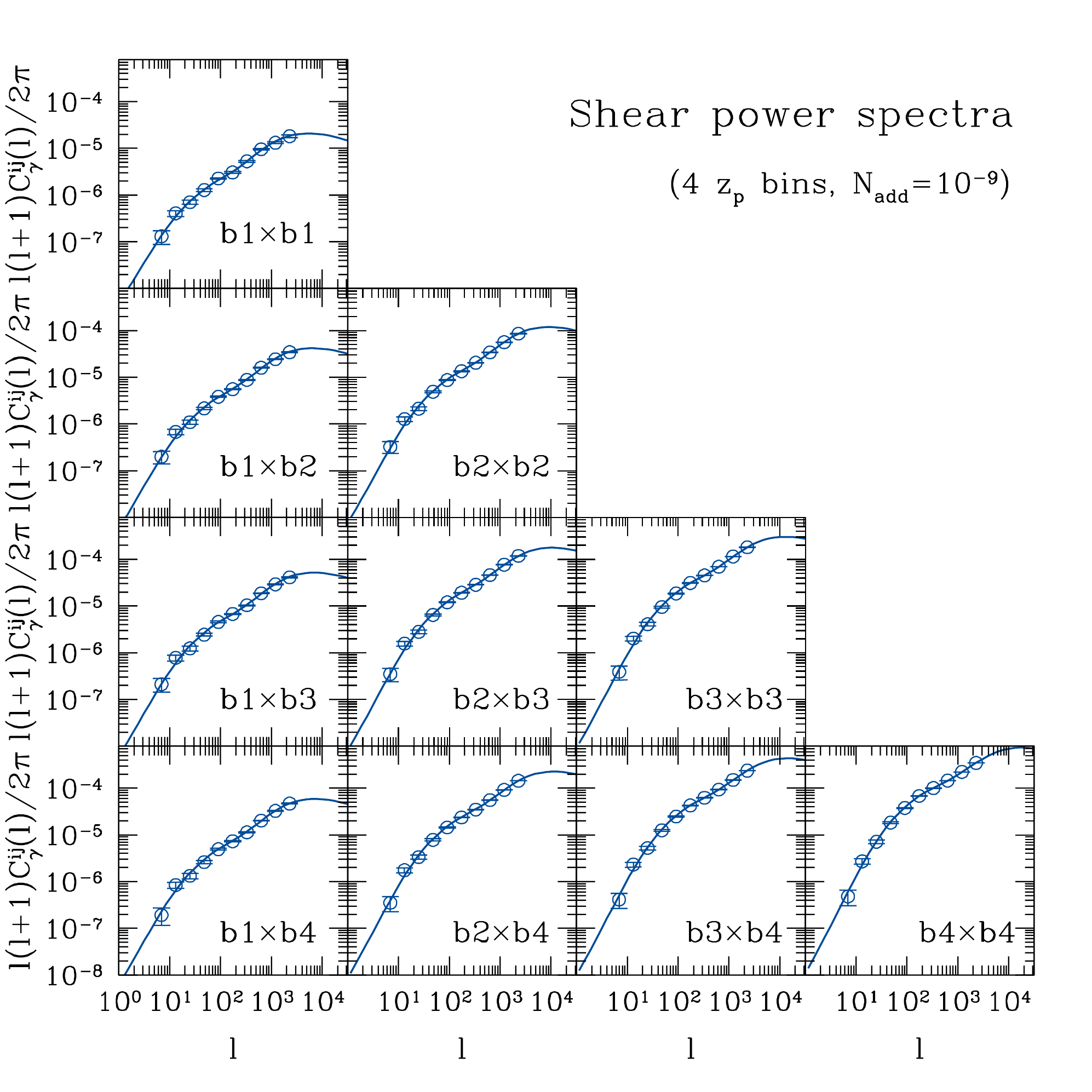}}
}
\caption{\label{fig:P_WL_4bins} The signal power spectra for the four tomographic bins assuming $\bar{N}_{\rm add}=10^{-9}$. Both auto and cross power spectra have been shown. To avoid the non-linear effect, we only consider the data points at $\ell<3000$ in the cosmological analysis.}
\end{figure*}

Considering intrinsic alignments and systematics, the measured shear power spectrum at a given multipole $\ell$ for the $i$th and $j$th tomographic bins can be estimated by \citep[e.g.][]{Huterer06,Amara08}
\be \label{eq:P_WL}
\widetilde{C}_{\gamma}^{ij}(\ell) = (1+m_i)(1+m_j)C_{\gamma}^{ij}(\ell) + \delta_{ij}\frac{\sigma_{\gamma}^2}{\bar{n}_i} + N^{ij}_{\rm add}(\ell),
\ee
where $C_{\gamma}^{ij}(\ell)$ is named as signal power spectrum, which is composed of three components \citep[e.g.][]{Hildebrandt16,Troxel17,Joudaki17}
\be
C_{\gamma}^{ij}(\ell) = P_{\kappa}^{ij}(\ell) + C_{\rm II}^{ij}(\ell) + C_{\rm GI}^{ij}(\ell).
\ee
Here $P_{\kappa}^{ij}(\ell)$ is the convergence power spectrum, which is the desired galaxy shear power spectrum for cosmological analysis. The $C_{\rm II}^{ij}(\ell)$ and $C_{\rm IG}^{ij}(\ell)$ are Intrinsic-Intrinsic (II), and Gravitational-Intrinsic (GI) power spectra, respectively. They are accounting for the intrinsic galaxy alignment effects, that ``II" denotes the correlation of the intrinsic ellipticities between neighbouring galaxies, and ``GI" means the correlation between the intrinsic ellipticity of a foreground galaxy and the gravitational shear of a background galaxy \citep[see e.g.][]{Joachimi16}. Assuming Limber and flat-sky approximations \citep{Limber54}, we have
\be
P_{\kappa}^{ij}(\ell) = \int_0^{\chi_{\rm H}} {\rm d}\chi\ \frac{q_i(\chi)q_j(\chi)}{r^2(\chi)} P_{\rm m}\left( \frac{\ell+1/2}{r(\chi)},\chi\right),
\ee
where $\chi$ is the comoving radial distance, $\chi_{\rm H}=\chi(z=5)$ denotes the horizon distance, and $r(\chi)$  is the comoving angular diameter distance. $P_{\rm m}$ is the matter power spectrum, which is calculated by the halo model \citep{Cooray02}. $q_i(\chi)$ is the lensing weighting function in the $i$th tomographic bin
\be
q_i(\chi) = \frac{3\Omega_{\rm m}H_0^2}{2c^2} \frac{r(\chi)}{a(\chi)} \int_{\chi}^{\chi_{\rm H}}{\rm d}\chi'\, n_i(\chi')\frac{r(\chi'-\chi)}{r(\chi')},
\ee
where $H_0=100\,h$ is the Hubble constant, $c$ is the speed of light, $a=1/(1+z)$ is the scale factor. $n_i(\chi)$ is the normalized source galaxy distribution of the $i$th tomographic bin, and $\int n_i(\chi){\rm d}\chi=1$. The intrinsic-intrinsic and gravitational-intrinsic power spectra are given by
\be
C_{\rm II}^{ij}(\ell) = \int_0^{\chi_{\rm H}} {\rm d}\chi \frac{n_i(\chi)n_j(\chi)F_i(\chi)F_j(\chi)}{r^2(\chi)} P_{\rm m}\left( \frac{\ell+1/2}{r(\chi)},\chi\right),
\ee
and
\ba
C_{\rm GI}^{ij}(\ell) &=& \int_0^{\chi_{\rm H}} {\rm d}\chi \frac{q_i(\chi)n_j(\chi)F_j(\chi)}{r^2(\chi)} P_{\rm m}\left( \frac{\ell+1/2}{r(\chi)},\chi\right) \nonumber \\ 
                               &+& \int_0^{\chi_{\rm H}} {\rm d}\chi \frac{q_j(\chi)n_i(\chi)F_i(\chi)}{r^2(\chi)} P_{\rm m}\left( \frac{\ell+1/2}{r(\chi)},\chi\right).
\ea
Here $F_i(\chi)$ is written as
\be
F_i(\chi) = -A_{\rm IA} C_1 \rho_{\rm c} \frac{\Omega_{\rm m}}{D(\chi)} \left( \frac{1+z}{1+z_0}\right)^{\eta_{\rm IA}} \left( \frac{L_i}{L_0}\right)^{\beta_{\rm IA}},
\ee
where $C_1=5\times10^{-14}\ h^{-2}M_{\sun}^{-1} {\rm Mpc}^3$, $\rho_{\rm c}$ is the present critical density, $D(\chi)$ is the linear growth factor normalized to unity at $z=0$, and $z_0=0.6$ and $L_0$ are pivot redshift and luminosity, respectively. $A_{\rm IA}$, $\eta_{\rm IA}$, and $\beta_{\rm IA}$ are free parameters in this model. For simplicity, we fix $\beta_{\rm IA}=0$ here, i.e. we do not consider luminosity dependence, since the change of the average luminosity can be ignored across different tomographic bins \citep{Hildebrandt16,Joudaki17}. The fiducial values of $A_{\rm IA}$ and $\eta_{\rm IA}$ are set to be -1 and 0, respectively, when producing mock data.

In the shot-noise term $N^{\rm shot}_{\gamma}$ of Eq.~(\ref{eq:P_WL}), $\sigma^2_{\gamma}=0.04$ is the shear variance per component caused by intrinsic ellipticity and measurement error, and $\bar{n}_i$ is the average galaxy number density in a given tomographic bin per steradian. According to the mock CSS-OS catalog, we assume $\sim100,000$ galaxies with $\sigma_z\le0.05$ per deg$^2$ (with $\sim 9.0\%$ outlier fraction) can be observed by the CSS-OS. This corresponds to a total density $\sim$28 arcmin$^{-2}$, and $\bar{n}_i=7.9$, 11.5, 4.6, and 3.7 $\rm arcmin^{-2}$ for the four redshift bins. Note that the number density estimated here can be smaller in the real observations, considering the complicated instrumental and astrophysical uncertainties. We need to perform realistic simulations to evaluate these effects in the future work.

Besides, we also consider the systematic errors in the CSS-OS weak lensing measurements. In Eq.~(\ref{eq:P_WL}), $m_i$ accounts for the effect of the multiplicative error in the $i$th redshift bin, which is averaged over all directions and galaxies in that bin. We assume it varies independently in different tomographic bins, and treat it as a free parameter in a given photo-$z$ bin in the fitting process \citep{Troxel17}. We set $m_i=0$ as fiducial value in each bin when generating mock data.

The $N^{ij}_{\rm add}$ in Eq.~(\ref{eq:P_WL}) is the additive error, which can be generated by the anisotropy of the PSF \citep{Huterer06}. In principle, $N^{ij}_{\rm add}(\ell)$ could vary at different scales and in different redshift bins, and also can appear in the correlations between bins \citep{Huterer06,Amara08,Amara10}. For simplicity, we adopt an average constant $\bar{N}_{\rm add}$ over all scales of all auto and cross shear power spectra for different tomographic bins \citep{Zhan06}. Based on the estimates of STEP (the Shear Testing Programme) and GREAT10 (the Gravitational Lensing Accuracy Testing 2010), the $\bar{N}_{\rm add}$ can be controlled within $10^{-10}$ with $S/N\sim10$ in the Stage IV weak lensing surveys \citep{Heymans06,Massey07,Kitching12,Massey13}. We find that the $S/N$ of most CSS-OS galaxies is around 7 for the most important $g$, $r$, and $i$ bands \citep{Cao18}. Although the $\bar{N}_{\rm add}$ are expected to achieve $\sim10^{-10}$ when adding up the flux of all seven CSST imaging bands, we would take a conservative estimate with $\bar{N}_{\rm add}=10^{-9}$ as a moderate value in this work. We will discuss $\bar{N}_{\rm add}=10^{-8}$ and $10^{-10}$ as pessimistic and optimistic cases.

The covariance matrix of the shear power spectra can be estimated by \citep{Hu04,Huterer06,Joachimi08}
\ba
&&{\rm Cov}\left[\widetilde{C}_{\gamma}^{ij}(\ell), \widetilde{C}_{\gamma}^{mn}(\ell')\right]  \nonumber \\
&=& \frac{\delta_{\ell\ell'}}{f_{\rm sky}\Delta l(2\ell+1)} \left[\widetilde{C}_{\gamma}^{im}(\ell) \widetilde{C}_{\gamma}^{jn}(\ell)+\widetilde{C}_{\gamma}^{in}(\ell) \widetilde{C}_{\gamma}^{jm}(\ell)\right],
\ea
where $\widetilde{C}^{ij}_{\gamma}(\ell)$ is the observed shear power spectrum given by Eq.~(\ref{eq:P_WL}), and $f_{\rm sky}$ is the sky coverage fraction of the survey. The CSS-OS can cover 17500 deg$^2$, but after removing masked area (covering image defects, reflections, ghosts, etc.), we assume an effective area of $\sim15000$ deg$^2$ can be used in the data analysis \citep{Hildebrandt16,Troxel17,Abbott17}.

In Figure~\ref{fig:P_WL}, we show the power spectra discussed above for the one photo-$z$ bin case (i.e. no tomographic bins). We assume that the shot-noise and additive terms can be eliminated in the data analysis\footnote{When fitting the real data, the shot-noise and additive terms can be fitted as free parameters in each tomographic bin in the fitting process. It could make the constraint results looser on the cosmological parameters than assuming eliminable case, since more parameters are included.}, and thus the mock data points of the $C_{\gamma}^{ij}$ can be derived as shown in blue circles with error bars. A random Gaussian distribution derived from the covariance matrix is added to each data point. The  $C_{\gamma}^{ij}$ for the four photo-$z$ bins are shown in Figure~\ref{fig:P_WL_4bins}. To avoid the non-linear effects, we only take account of the data at $\ell<3000$.

\section{Galaxy clustering survey}

\begin{figure}[t]
\includegraphics[scale = 0.42]{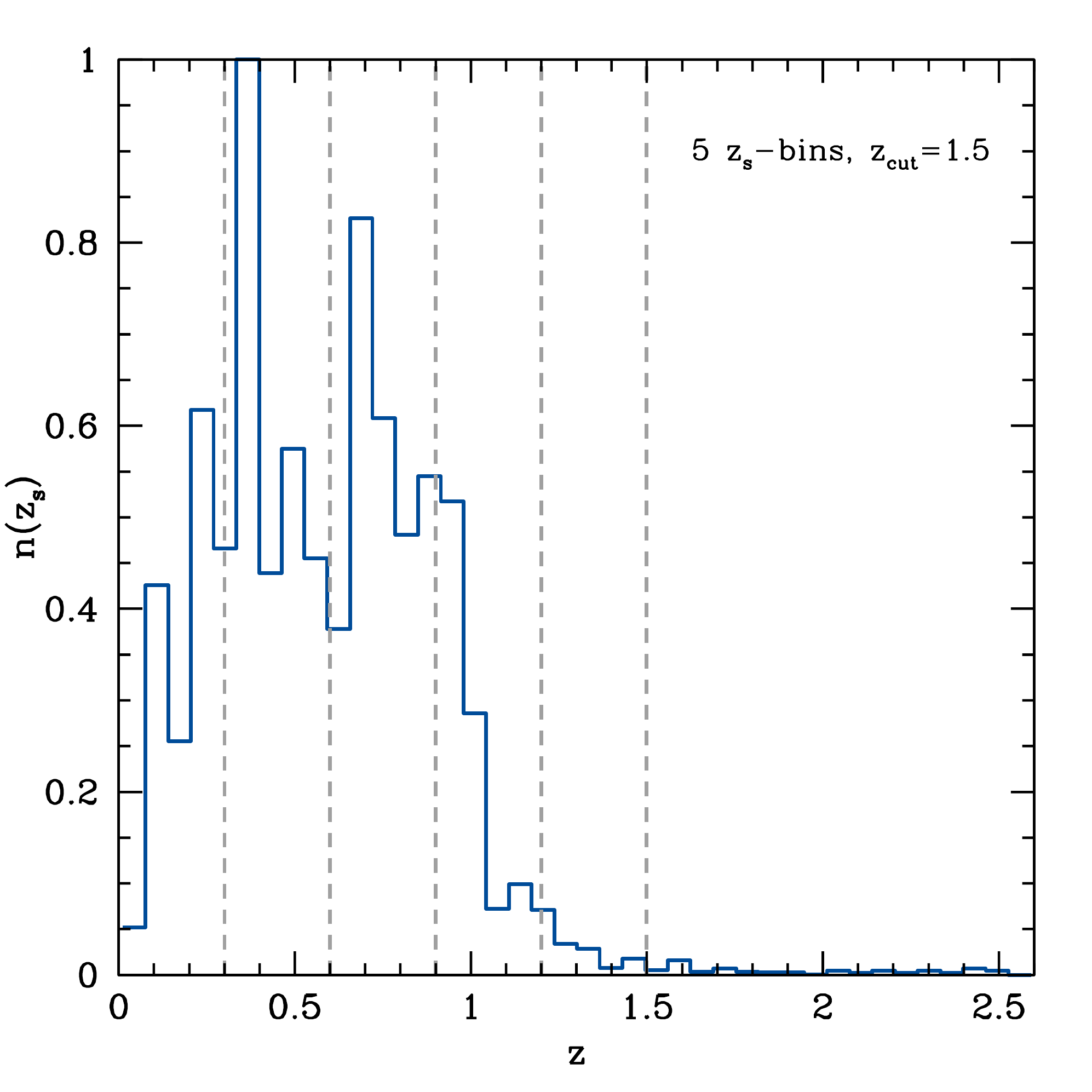}
\caption{\label{fig:nz_spec} The mock galaxy redshift distribution of the CSST spectroscopic survey. The zCOSMOS catalog, which has similar survey depth, is adopted to simulate the CSST survey result. We find that the distribution has a peak around $z=0.3-0.4$, and can reach up to $z\sim2.5$. We divide the redshift range into five bins (by the gray vertical lines) with $\Delta_z=0.3$ for tomographic study.}
\end{figure}

\begin{figure*}
\centerline{
\resizebox{!}{!}{\includegraphics[scale=0.8]{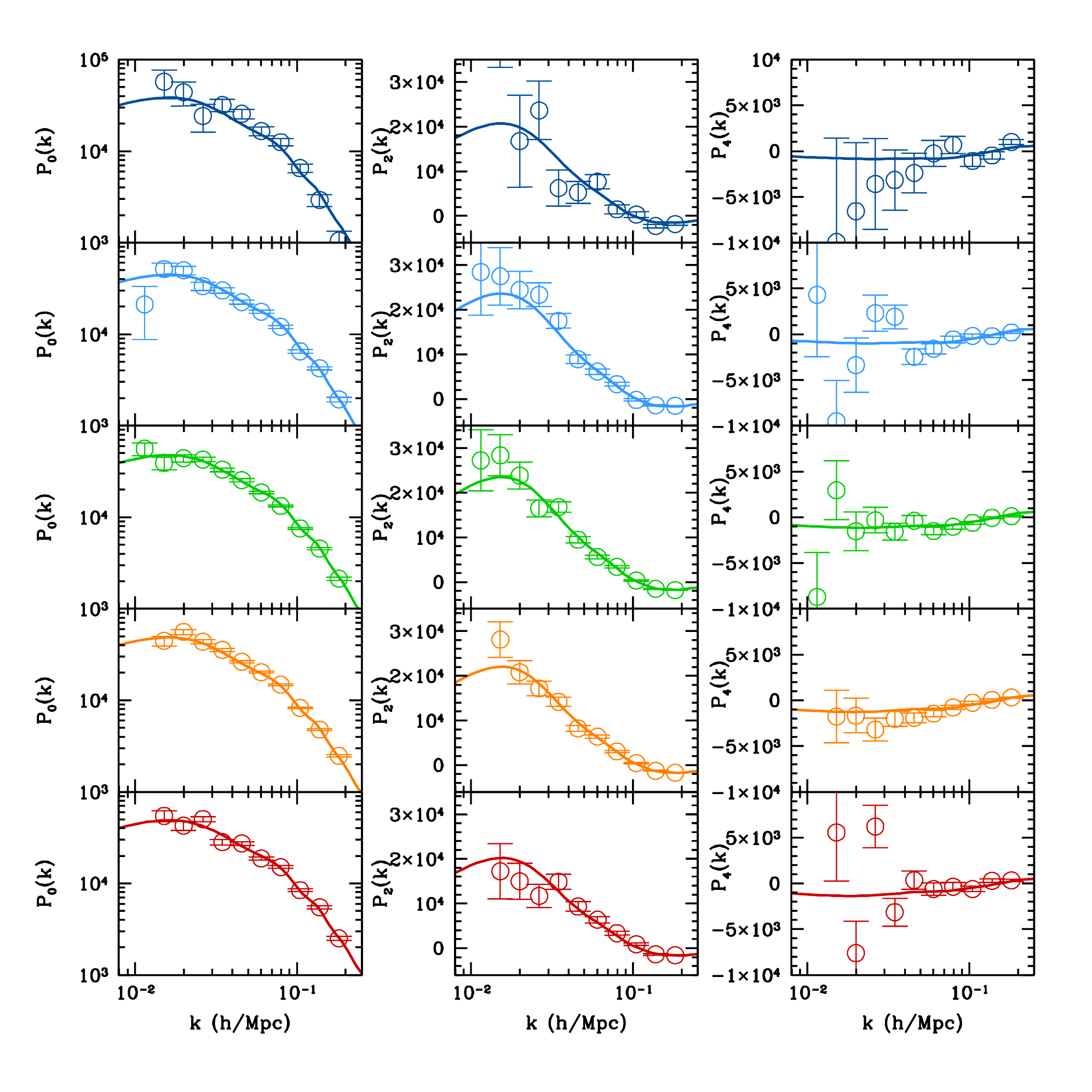}}
}
\caption{\label{fig:P_RSD_5bins} The mock galaxy multipole moments of power spectra $P_0^g$, $P_2^g$, and $P_4^g$. The multipole power spectra for the five spec-$z$ bins (from low to high redshift) are shown from the top to the bottom panels. We only consider the $P^g_{\ell}$ at $k<0.2$ $h/{\rm Mpc}$ to avoid the non-linear clustering effect. We assume $f^{z_{\rm s},0}_{\rm eff}=0.5$ and $N^g_{\rm sys}=5\times10^4$ $({\rm Mpc}/h)^3$ as a moderate case when generating the error bars.}
\end{figure*}

In addition to the photometric imaging survey, the CSS-OS also can simultaneously perform the spectroscopic survey using slitless gratings. The CSST spectroscopic survey covers the same survey area (17500 deg$^2$) and similar  wavelength range (255-1000 nm) as the photometric imaging survey. It contains three bands, i.e. $GU$, $GV$, and $GI$ (see Figure \ref{fig:filters}), with AB magnitude $5\sigma$ limit $\sim$21 mag per resolution element for point sources with spectral resolution $R\ge200$ (see Table \ref{tab:Des_paramts}). In this section, we will discuss the galaxy clustering power spectrum with the effect of redshift-space distortion (RSD) measured by the CSST spectroscopic survey.

\subsection{Galaxy spectroscopic redshift distribution}

We adopt the zCOSMOS catalog (DR3 release) to simulate the CSST spectroscopic survey result. The zCOSMOS  redshift survey observes in the COSMOS field using the VIMOS spectrograph mounted at the Melipal Unit Telescope of the VLT \citep{Lilly07,Lilly09}. It covers 1.7 deg$^2$ with a magnitude limit $I_{\rm AB}\simeq22.5$, which is close to survey depth of the CSS-OS. The spectral coverage is 5550-9450 $\rm \AA$ with a spectral resolution $R\sim600$. There are about 20,000 sources in this catalog, and after selecting high-quality data suggested by the zCOSMOS team, we obtain about 16,600 sources (80\% of the total) with reliable spectroscopic redshifts.

The redshift distribution of this mock CSST spectroscopic catalog is shown in Figure~\ref{fig:nz_spec}. We can see that it has a peak at $z=0.3-0.4$, and can extend to $z\sim2.5$. Since there are not many galaxies at high redshifts, we only consider the sources at $z<1.5$ in the galaxy clustering analysis ($\sim40$ galaxies are out of this range). In order to study the evolution of the equation of state of dark energy and other cosmological parameters, we divide the redshift range into five tomographic bins with $\Delta_z=0.3$ assuming no spec-$z$ outliers for simplicity. The galaxy number fraction in the five spectroscopic bins are 0.17, 0.37, 0.35, 0.10, and 0.01, respectively. Note that more number of redshift bins can be used in the real data analysis, depending on the number density of observed galaxies.

\subsection{Redshift-space galaxy power spectrum}

The redshift-space galaxy power spectrum in ($k$, $\mu$) dimensions can be measured by the CSST spectroscopic survey, which can be expanded in Legendre polynomials \citep{Taylor96,Ballinger96}
\be
P^{\rm (s)}_{g}(k,\mu) = \sum_{\ell} P^g_{\ell}(k)\, \mathcal{L}_{\ell}(\mu),
\ee
where (s) denotes the redshift space, $\mu=k_{\parallel}/k$ is the cosine of the angle between the direction of wavenumber $k$ and the line of sight, $\mathcal{L}_{\ell}(\mu)$ is the Legendre polynomials that only the first few non-vanishing orders $\ell=(0, 2, 4)$ are considered in the linear regime, and $P^g_{\ell}(k)$ is the multipole moments of the power spectrum. Considering the Alcock-Paczynski effect \citep{Alcock79}, the galaxy multipole power spectra are given by
\be
P^g_{\ell}(k) = \frac{2\ell+1}{2\alpha^2_{\perp}\alpha_{\parallel}} \int_{-1}^1 {\rm d}\mu\, P^{\rm (s)}_{g}(k',\mu') \mathcal{L}_{\ell}(\mu).
\ee
Here $\alpha_{\perp}=D_{\rm A}(z)/D^{\rm fid}_{\rm A}(z)$ and $\alpha_{\parallel}=H^{\rm fid}(z)/H(z)$ are the scaling factors in the transverse and radial directions, respectively, and the superscript ``fid" means the quantities in the fiducial cosmology. The $k'=\sqrt{k'^2_{\parallel}+k'^2_{\perp}}$ and $\mu'=k'_{\parallel}/k'$ are the apparent wavenumber and cosine of angle, where $k'_{\parallel}=k_{\parallel}/\alpha_{\parallel}$ and $k'_{\perp}=k_{\perp}/\alpha_{\perp}$. Assuming there is no peculiar velocity bias, the apparent redshift-space galaxy power spectrum can be written as
\be
P^{\rm (s)}_{g}(k',\mu') = P_g(k') (1+\beta\mu'^2)^2\, \mathcal{D}(k',\mu'),
\ee
where $P_g(k')=b_{g}^2\, P_{\rm m}(k')$ is the apparent real-space galaxy power spectrum, $b_{g}$ is the galaxy bias, $P_{\rm m}$ is the matter power spectrum, and $\beta=f/b_{g}$ where $f={\rm d\,ln}D(a)/{\rm d\, ln}\,a$ is the growth rate. $\mathcal{D}(k', \mu')$ is the damping term at small scales, which is given by
\be
\mathcal{D}(k',\mu') = {\rm exp}\left[ -\left(k'\mu'\sigma_{\rm D}\right)^2\right].
\ee
Here $\sigma^2_{\rm D}=\sigma_v^2+\sigma_R^2$, where $\sigma_v=\sigma_{v0}/(1+z)$ is the velocity dispersion \citep{Scoccimarro04,Taruya10}, and we assume $\sigma_{v0}=7$ ${\rm Mpc}/h$ for the measurements of emission line galaxies in the CSS-OS \citep{Blake16,Joudaki17}. It will be set as a free parameter in the fitting process. The $\sigma_R=c\,\sigma_z/H(z)$ is the smearing factor for the small scales below the spectral resolution of spectroscopic surveys, where $H(z)$ is the Hubble parameter, and $\sigma_z=(1+z)\sigma^0_{z}$ \citep{Wang09}. Based on the instrumental design of the CSST, we set $\sigma^0_{z}=0.002$ as the accuracy of the spectral calibration. Note that this accuracy could be larger in the real observation, and we need to run simulations in the future for further confirmation.

After adding the shot-noise term and systematics, we obtain the total multipole power spectra of the $a$th spec-$z$ bin
\be
\widetilde{P}^{g,\,a}_{\ell}(k) = P^{g,\,a}_{\ell}(k) + \frac{1}{\bar{n}_g^a} + N^{g,\,a}_{{\rm sys}},
\ee
where $\bar{n}_g^a$ is the galaxy number density in the $a$th spec-$z$ bin. We take an average value $\bar{n}_g^a$ in a given redshift bin, and find that $\bar{n}_{g,\rm ori}^{a}=3.4\times10^{-2}$, $1.1\times10^{-2}$, $5.5\times10^{-3}$, $1.2\times10^{-3}$ $({\rm Mpc}/h)^{-3}$, and $7.9\times10^{-5}$ from the mock CSS-OS catalog for the five spec-$z$ bins, respectively. Besides, we consider an effective redshift factor $f^{z_s}_{\rm eff}$ to account for the fraction of galaxies that can achieve the required accuracy of the CSS-OS spec-$z$ calibration with $\sigma^0_{z}=0.002$ in the slitless grating observations \citep{Wang10}. We assume the fraction decreases as the redshift increases, i.e. $f^{z_{\rm s}}_{\rm eff}=f^{z_{\rm s},0}_{\rm eff}/(1+z)$, where we set $f^{z_{\rm s},0}_{\rm eff}=0.3$, 0.5, and 0.7, respectively, as the pessimistic, moderate, and optimistic cases. So the final number density in the $a$th spec-$z$ bin should be $\bar{n}_g^a=f^{z_s,a}_{{\rm eff}}\,\bar{n}_{g,\rm ori}^{a}$. The systematics $N^g_{{\rm sys}}$ due to instrumentation effects of the CSST slitless gratings are assumed to be a constant in different tomographic bins, or it can be seen as the average value  $\bar{N}^g_{\rm sys}$ over all redshift bins and scales. Combining the three assumptions of $f^{z_{\rm s},0}_{\rm eff}$, we jointly assume three values, i.e. $\bar{N}^g_{\rm sys}=10^{5}$, $5\times10^{4}$, and $10^{4}$ $({\rm Mpc}/h)^3$, as pessimistic, moderate, and optimistic cases in the CSST galaxy clustering surveys.

\begin{figure}[t]
\includegraphics[scale = 0.42]{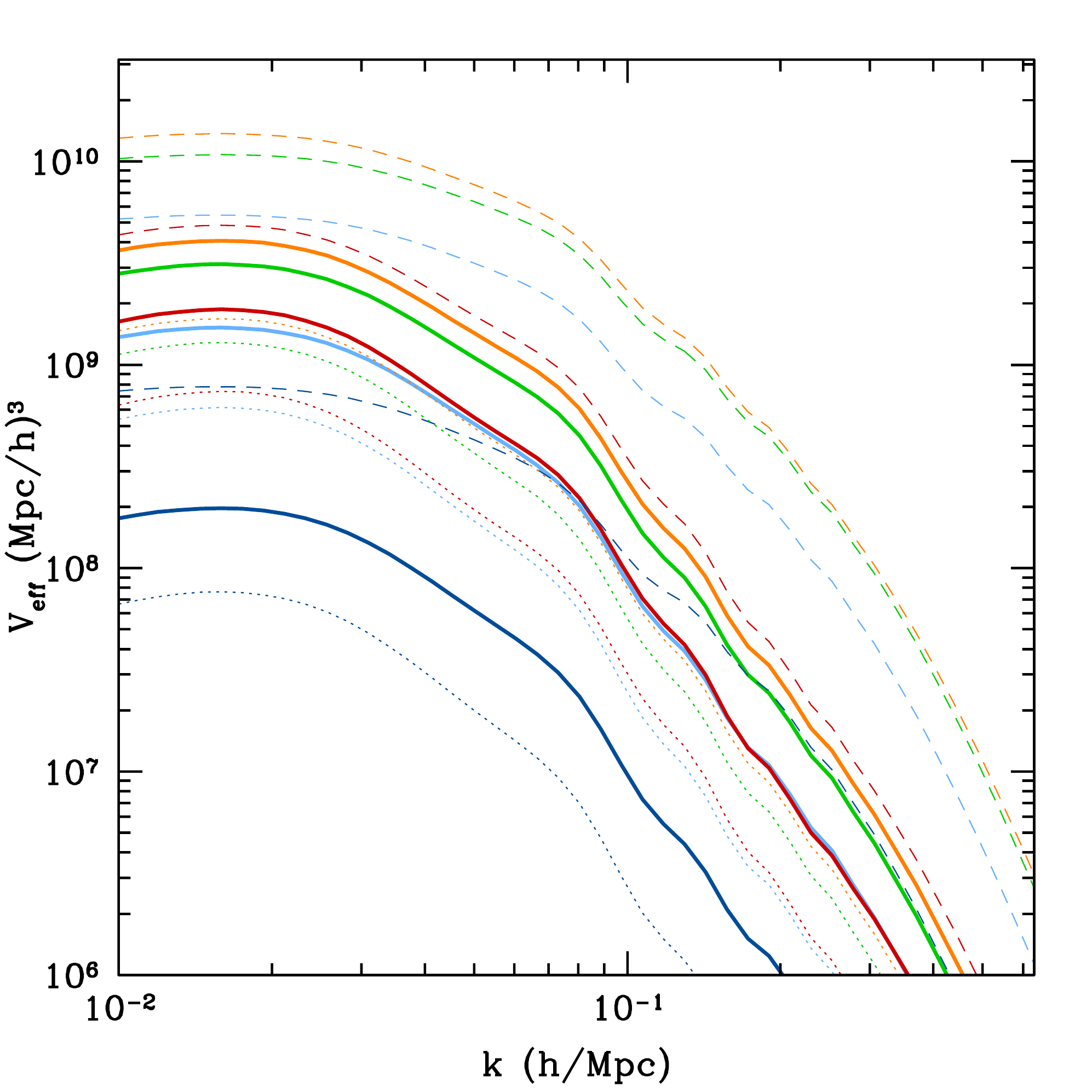}
\caption{\label{fig:Veff} The $V_{\rm eff}$ of the $P_0^g$ for the five spectroscopic redshift bins. The blue, light blue, green, orange, and red curves are for the 1st, 2nd, 3rd, 4th, and 5th spec-$z$ bin, respectively. The dotted, solid, and dashed curves denote the results for the pessimistic, moderate, and optimistic cases with $f^{z_{\rm s},0}_{\rm eff}=0.3$, 0.5, 0.7, and $\bar{N}^g_{\rm sys}=10^5$, $5\times10^4$, $10^4$ $({\rm Mpc}/h)^3$, respectively.}
\end{figure}

The errors of the multipole power spectra in the $a$th bin are then given by
\be
\sigma^a_{P^g_{\ell}}(k) = 2\pi\sqrt{\frac{1}{V^a_{\rm S}\,k^2\Delta k}}\ \widetilde{P}^{g,\,a}_{\ell}(k),
\ee
where $V^a_{\rm S}$ is the survey volume of the $a$th bin. We assume a $15000$ deg$^2$ sky coverage for the galaxy clustering spectroscopic survey, which is the same as the photometric imaging survey area used in the weak lensing, after masking the area with bad measurements. For a constant $n_g$ in a given redshift bin, the corresponding effective survey volume for a multipole component of power spectrum can be defined as
\be
V^{\ell,\, a}_{\rm eff}(k) \equiv \left[1 + \frac{1}{\bar{n}_g^aP^{g,\,a}_{\ell}(k)} + \frac{N^{g,\,a}_{\rm sys}}{P^{g,\,a}_{\ell}(k)}\right]^{-2} V^a_{\rm S}.
\ee

In Figure~\ref{fig:P_RSD_5bins}, we show the mock multipole power spectra $P_0^g$, $P_2^g$, and $P_4^g$. In order to avoid the non-linear effect, we only consider the data at $k<0.2$ $h/{\rm Mpc}$. The errors of  the mock data are generated by assuming $f^{z_s,0}_{\rm eff}=0.5$ and $\bar{N}^g_{\rm sys}=5\times10^4$ $({\rm Mpc}/h)^3$. In Figure~\ref{fig:Veff}, the $V_{\rm eff}$ of the $P_0^g$ for each redshift bin are shown. The dashed, solid, and dotted curves are for the optimistic, moderate, and pessimistic cases.

In addition to the redshift-space power spectrum, the CSS-OS three-dimensional (3-d) galaxy clustering data  also can be analyzed using other methods, such as topology \citep{topology}, tomographic Alcock-Paczynski \citep{Li2016} and 3-point correlation function \citep{Takada:2002qq}, that can extract more cosmological information. We will discuss these methods in the future work.

\section{Galaxy-galaxy lensing survey}

We can also cross-correlate the WL and galaxy clustering surveys to get galaxy-galaxy lensing power spectra for the photo-$z$ and spec-$z$ bins in the CSS-OS. This can help us to derive more cosmological information.

\subsection{Angular galaxy power spectrum}

We first consider the two-dimensional (2-d) angular galaxy power spectrum for the $a$th and $b$th spec-$z$ bins, which can be derived from the 3-d galaxy power spectrum by assuming Limber approximation \citep{Limber54,Kaiser92,Hu04}
\be
C_{g}^{ab}(\ell) = \int_0^{\chi_{\rm H}} {\rm d}\chi\ \frac{b^2_g(k,\chi)\hat{n}^a_g(\chi)\hat{n}^b_g(\chi)}{r^2(\chi)} P_{\rm m}\left( \frac{\ell+1/2}{r(\chi)},\chi\right),
\ee
where $b_g(k,\chi)$ is the galaxy bias, which depends on the scales and redshifts. For simplicity, we assume it is a constant at different scales and only varies as a function of redshift $b_g(z)$ when producing mock data. $\hat{n}^a_g(\chi)$ is the normalized galaxy redshift distribution for the $a$th spec-$z$ bin, that we have $\int \hat{n}^a_g(\chi){\rm d}\chi=1$. Incorporating the shot-noise term and systematics, the total angular galaxy power spectrum can be written as
\be \label{eq:C_gal_obs}
\widetilde{C}_{g}^{ab}(\ell) = C_{g}^{ab}(\ell) + \frac{\delta_{ab}}{\bar{n}'^a_g} + N'^{\,g, \,ab}_{\rm sys}(\ell),
\ee
where $\bar{n}'^a_g$ is the average galaxy number density in the $a$th spec-$z$ bin per steradian, and it is found to be $0.46$, $1.0$, $0.94$, $0.29$, and $0.02$ arcmin$^{-2}$ for the five bins ($\sim$ 2.7 arcmin$^{-2}$ totaly), respectively. $N'^{\,g, \,ab}_{\rm sys}(\ell)$ is the systematic noise for the angular galaxy power spectrum, and we assume it is a constant for different scales and redshift bins. We find that the result is not sensitive to the $N'^{\,g}_{\rm sys}$ as long as it is less than $10^{-8}$, i.e. the shot-noise term (the second term) in Eq.~(\ref{eq:C_gal_obs}) is relatively large and dominant.

\begin{figure}[t]
\includegraphics[scale = 0.42]{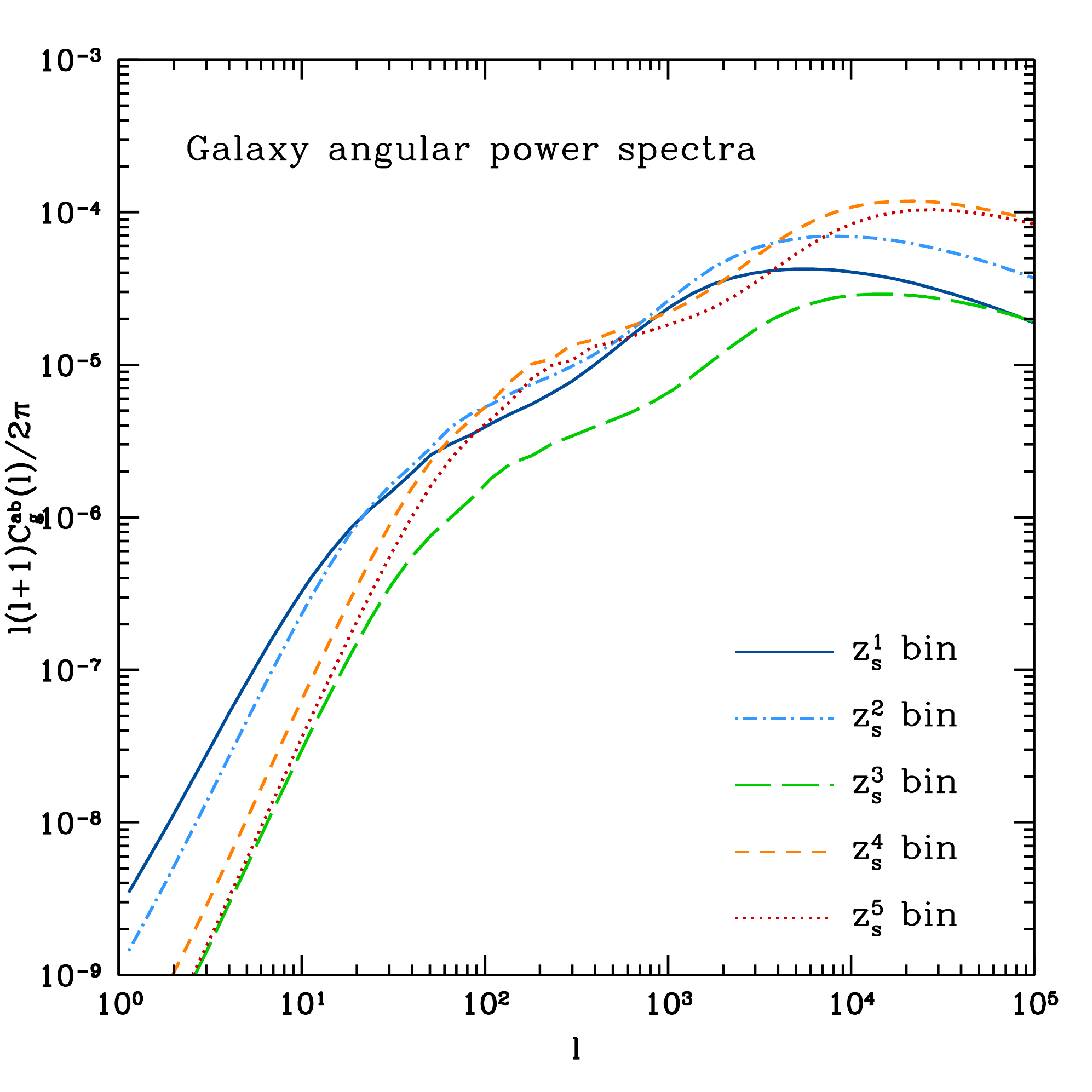}
\caption{\label{fig:P_gal_2D} The angular galaxy power spectra for the five spectroscopic redshift bins, which can be cross-correlate with the shear power spectra. Note that there is no cross power spectrum between spec-$z$ bins. }
\end{figure}

Note that the cross power spectra vanish in the spectroscopic surveys (e.g. the case we discuss here), since there is no overlapping region between spec-$z$ bins. However, for the photometric surveys, both the auto and cross angular galaxy power spectra are important, and can be used to calibrate the galaxy bias and redshift distributions when cooperating with the weak lensing survey \citep{Hu04,Zhan06}. The CSS-OS also can probe the angular galaxy power spectrum in the photometric imaging survey, and we will discuss it in our future work. In Figure~\ref{fig:P_gal_2D}, we show the 2-d angular galaxy power spectra for the five spec-$z$ bins. 

\subsection{Galaxy-galaxy lensing power spectrum}

Then we can calculate the angular galaxy-galaxy lensing power spectrum, i.e. cross-correlating the galaxy clustering and weak lensing surveys, of the $a$th galaxy spec-$z$ bin and the $i$th photo-$z$ bin. Considering the intrinsic alignment effect, it can be expressed as \citep{Joudaki17}
\be
C_{g\gamma}^{ai}(\ell) = C_{g\kappa}^{ai}(\ell) + C_{g{\rm I}}^{ai}(\ell),
\ee
where the $C_{g\kappa}^{ai}$ and $C_{g{\rm I}}^{ai}$ are given by
\be
C_{g\kappa}^{ai}(\ell) = \int_0^{\chi_{\rm H}} {\rm d}\chi\ \frac{b_g(k,\chi)\hat{n}^a_g(\chi)q_i(\chi)}{r^2(\chi)} P_{\rm m}\left( \frac{\ell+1/2}{r(\chi)},\chi\right),
\ee
and
\be
C_{g{\rm I}}^{ai}(\ell) = \int_0^{\chi_{\rm H}} {\rm d}\chi\ \frac{b_g(k,\chi)\hat{n}^a_g(\chi)n_i(\chi)F_i(\chi)}{r^2(\chi)} P_{\rm m}\left( \frac{\ell+1/2}{r(\chi)},\chi\right).
\ee
The covariance matrix for the galaxy-galaxy lensing power spectra is given by \citep{Hu04}
\ba
&&{\rm Cov}\left[C_{g\gamma}^{ai}(\ell), C_{g\gamma}^{bj}(\ell')\right]  \nonumber \\
&=& \frac{\delta_{\ell\ell'}}{f_{\rm sky}\Delta l(2\ell+1)} \left[\widetilde{C}_{g}^{ab}(\ell) C_{\gamma}^{ij}(\ell)+C_{g\gamma}^{aj}(\ell) C_{g\gamma}^{ib}(\ell)\right].
\ea

\begin{figure}[t]
\includegraphics[scale = 0.42]{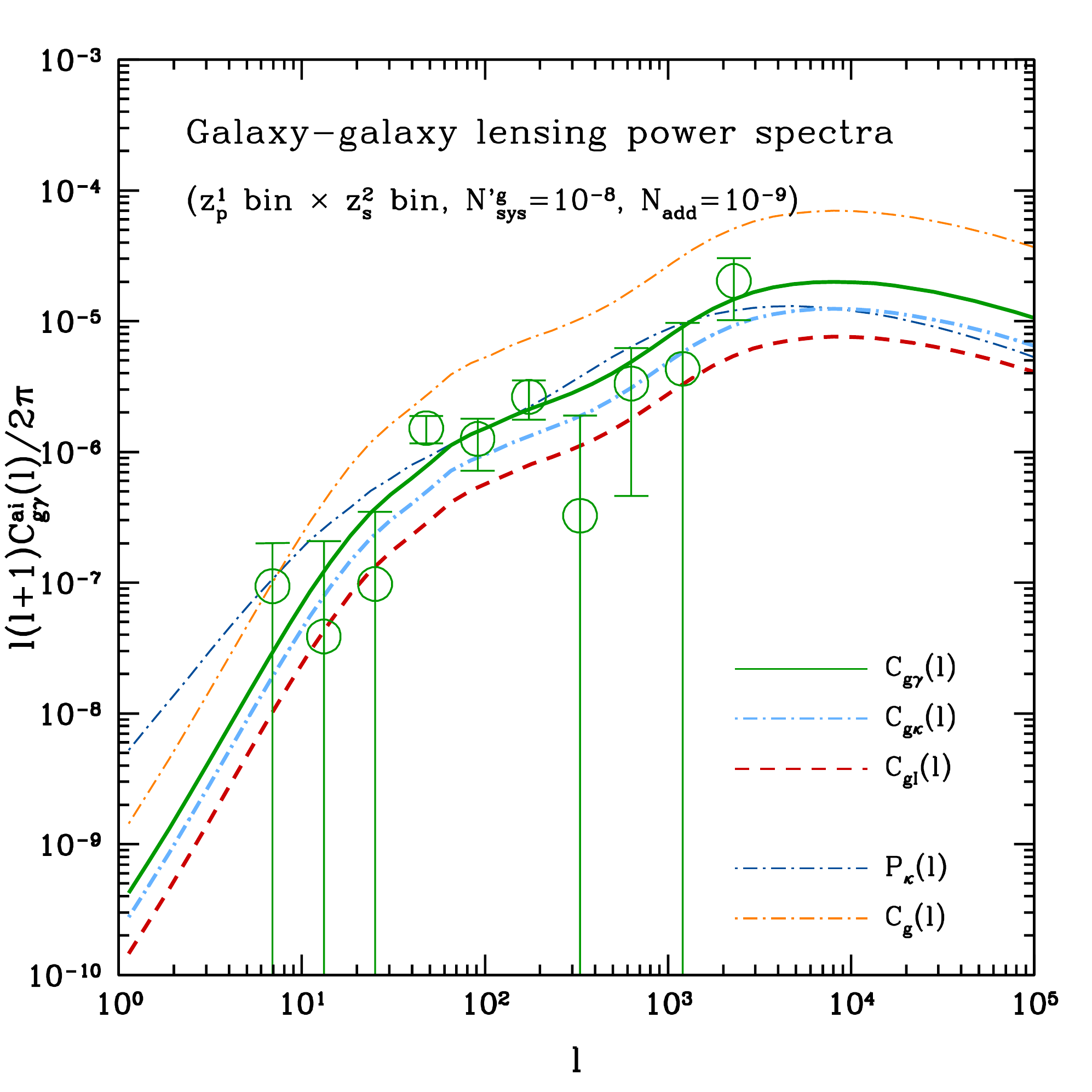}
\caption{\label{fig:P_WLxgal} The angular galaxy-galaxy lensing power spectra for the first photometric and second spectroscopic redshift bins, considering the intrinsic alignment effect. For comparison, the corresponding convergence and angular galaxy power spectra are also shown.}
\end{figure}

\begin{figure}[t]
\includegraphics[scale = 0.42]{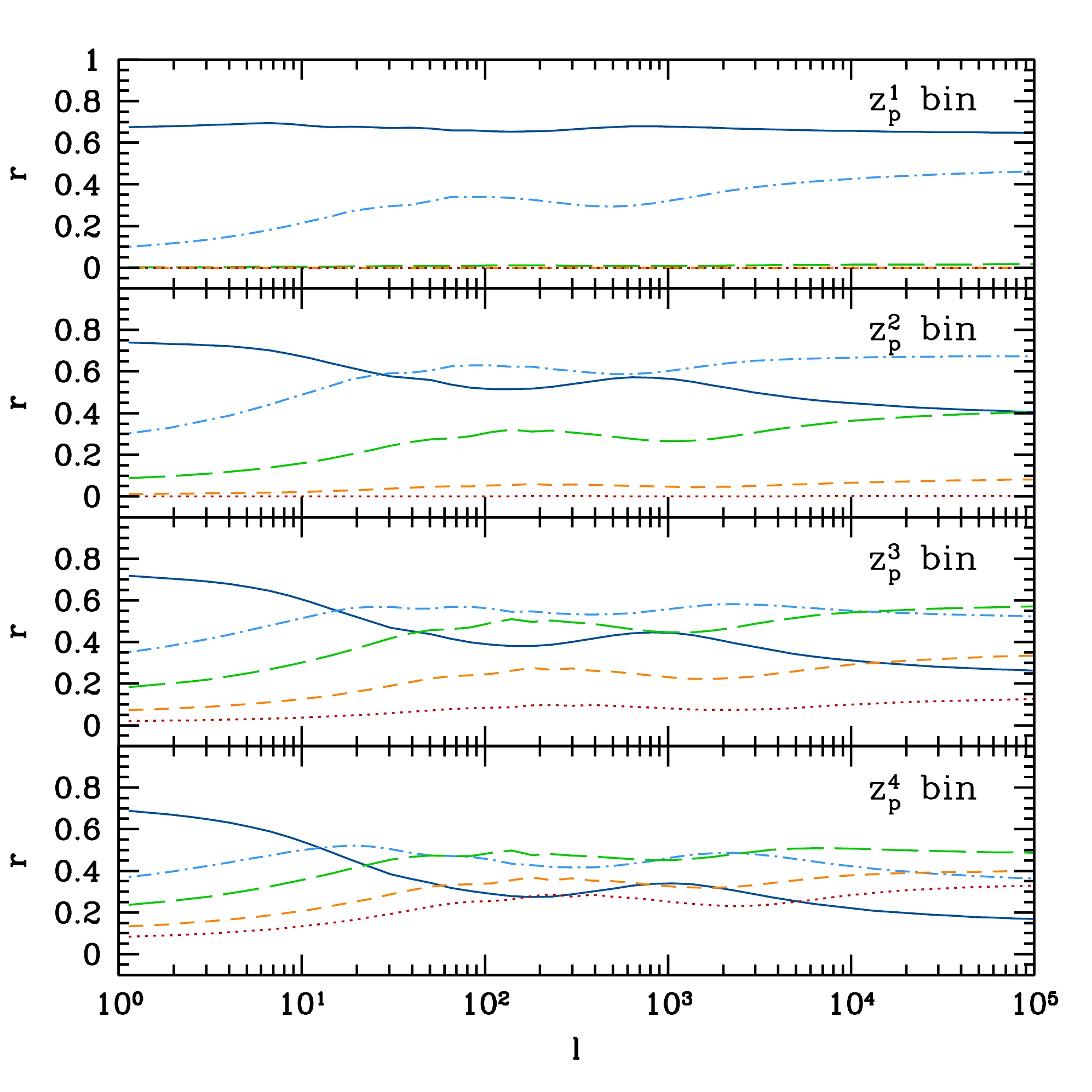}
\caption{\label{fig:r} The coefficients of the cross power spectra for the four photometric and five spectroscopic bins. The solid blue, light blue dash-dotted, green long dashed, orange dashed, and red dotted curves are for the correlations between a given photo-$z$ bin (from the top to bottom panels) and the 1st, 2nd, 3rd, 4th, and 5th spec-$z$ bins, respectively.}
\end{figure}

In Figure~\ref{fig:P_WLxgal}, we show the mock galaxy-galaxy lensing power spectra for the first photo-$z$ and second spec-$z$ bins. The green solid, light blue dash-dotted, red dashed curves denote the total, galaxy-galaxy lensing, and galaxy-Intrinsic power spectra, respectively. For comparison, the corresponding convergence (dark blue dash-dotted) and angular galaxy (orange dashed-dotted) power spectra are also shown. The data points with error bars are obtained by assuming $\bar{N}_{\rm add}=10^{-9}$ and $N'^{\,g}_{\rm sys}=10^{-8}$.

We can define the coefficients of the cross power spectra of weak lensing and galaxy clustering between the photometric and spectroscopic redshift bins, which is given by
\be
r_{ai}(\ell) = \frac{C^{ai}_{g\kappa}(\ell)}{\sqrt{C^{aa}_g(\ell)\,P^{ii}_{\kappa}(\ell)}}.
\ee
In Figure~\ref{fig:r}, we calculate the coefficients of the cross power spectra for the four photometric and five spectroscopic bins. For instance, in the top panel, we find that the $z_{\rm p}^1$ bin only have significant correlations with the $z_{\rm s}^1$ and $z_{\rm s}^2$ bins, since they cover similar redshift range (see Figure~\ref{fig:nz_phot} and \ref{fig:nz_spec}). 

\section{Fitting the mock data}

When generating the mock data, we assume the flat $\Lambda$CDM cosmology with the equation of state (EoS) of dark energy $w=-1$. In order to explore the constraint of the CSS-OS on the evolution of the dark energy EoS, we adopt the $w$CDM model with $w=w_0+w_a(1-a)$ when fitting the mock data. The {\tt HMcode} is used to calculate the non-linear matter power spectrum for the $w$CDM model \citep{Mead15,Mead16}. The cosmological and systematical parameters for the weak lensing, galaxy clustering, and galaxy-galaxy lensing power spectra in the fitting process are shown in Table~\ref{tab:parameters}. In the joint constraint case of the CSST WL and galaxy clustering surveys with four photo-$z$ bins and five spec-$z$ bins, we totally have 31 free parameters in the model.

\begin{table}
\centering
\caption{The fiducial values and ranges of the free parameters in the model.}
\label{tab:parameters}
\begin{tabular}{ c  c  c  }
\hline\hline\vspace{0.1cm}
Parameter & fiducial value & flat prior  \\
\hline
\vspace{0.1cm}
& Cosmology &\\
$\Omega_{\rm m}$ & 0.3 & (0, 1)\\
$\Omega_{\rm b}$ & 0.05 & (0, 0.1)\\
$\sigma_8$ & 0.8 & (0.4, 1) \\
$n_{\rm s}$ & 0.96 & (0.9, 1) \\
$w_0$ & -1 & (-5, 5) \\
$w_a$ & 0 & (-10, 10) \\
$h$ & 0.7 & (0, 1)\vspace{0.1cm} \\
\hline
\vspace{0.1cm}
& Intrinsic alignment & \\
$A_{\rm IA}$ & -1 & (-5, 5) \\
$\eta_{\rm IA}$ & 0 & (-5, 5) \vspace{0.1cm}\\
\hline\vspace{0.1cm}
& Galaxy bias &\\
$b_g^a$ & (1.15, 1.45, 1.75, 2.05, 2.35) & (0, 4)\vspace{0.1cm} \\
\hline\vspace{0.1cm}
& Velocity dispersion &\\
$\sigma_{v0}^a$ & (6.09, 4.83, 4.0, 3.41, 2.98) & (0, 10)\vspace{0.1cm} \\
\hline\vspace{0.1cm}
& Photo-$z$ calibration &\\
${\Delta z}_i$  & (0, 0, 0, 0) & (-0.1, 0.1) \\
${\rm log_{10}}(s_z^i)$ & (0, 0, 0, 0) & (-0.1, 0.1) \vspace{0.1cm}\\
\hline\vspace{0.1cm}
& Shear calibration &\\
$m_i$ & (0, 0, 0, 0) & (-0.1, 0.1) \\
\hline
\end{tabular}
\end{table}

\begin{figure*}[t]
\centerline{
\includegraphics[scale=0.34]{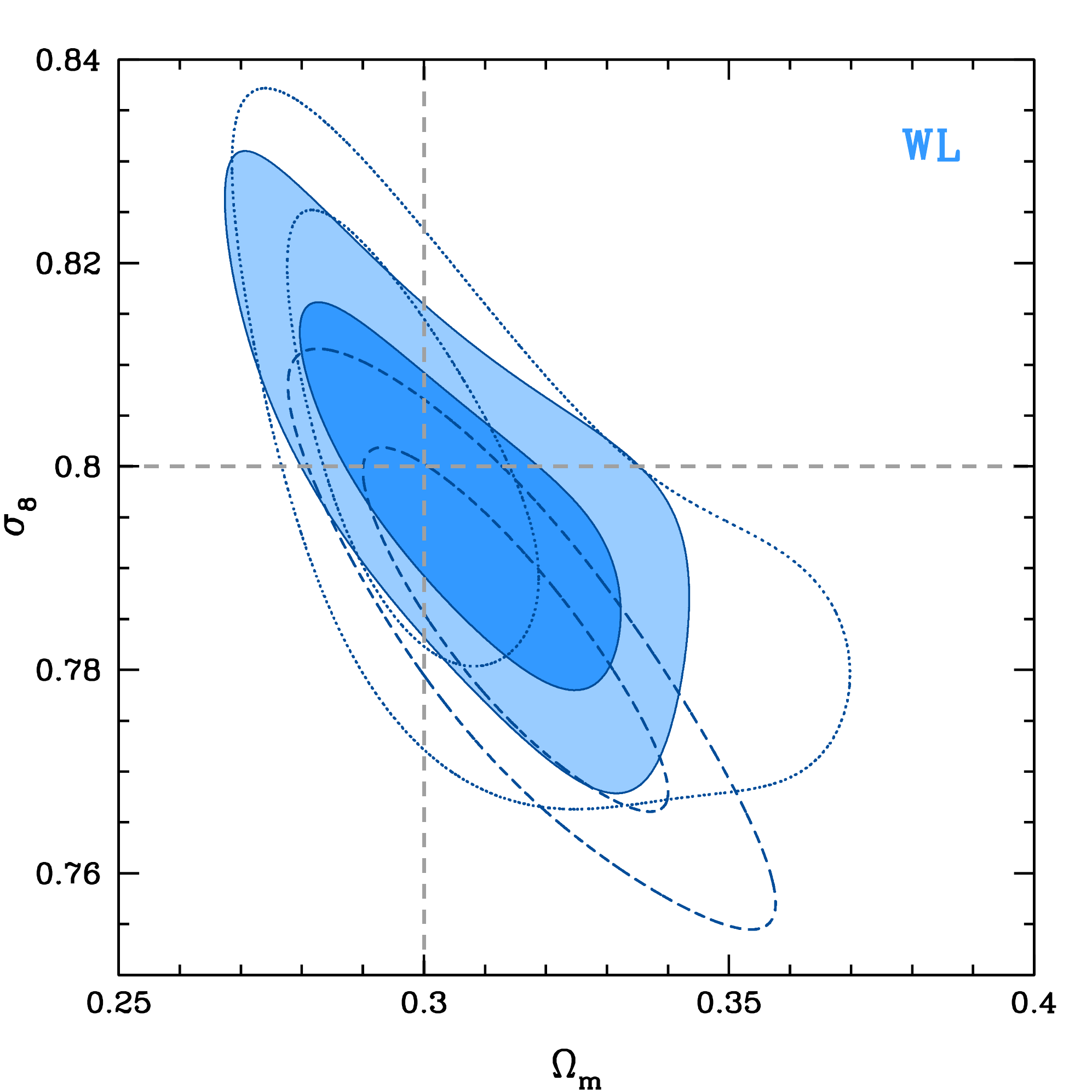}
\includegraphics[scale=0.34]{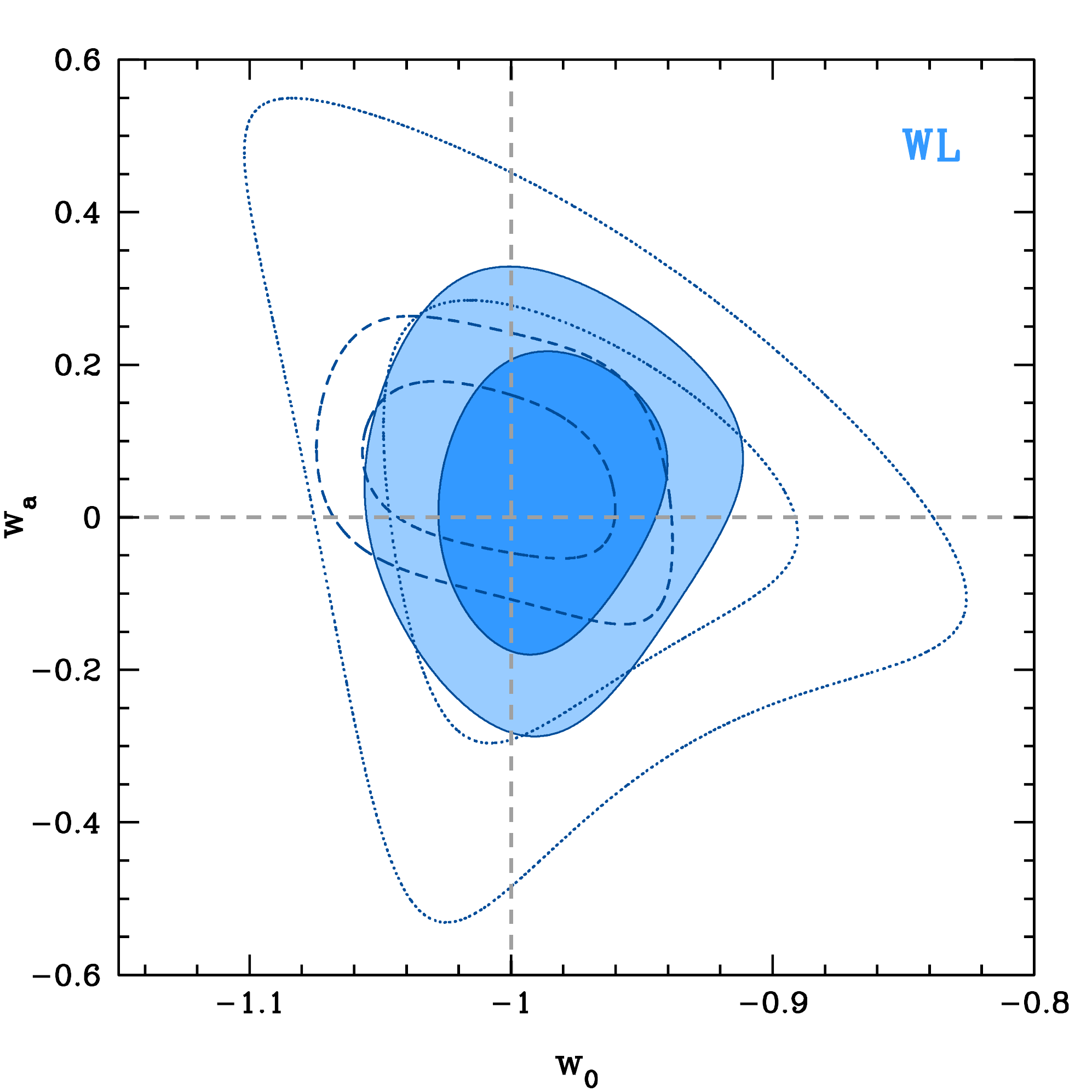}}
\centerline{
\includegraphics[scale=0.34]{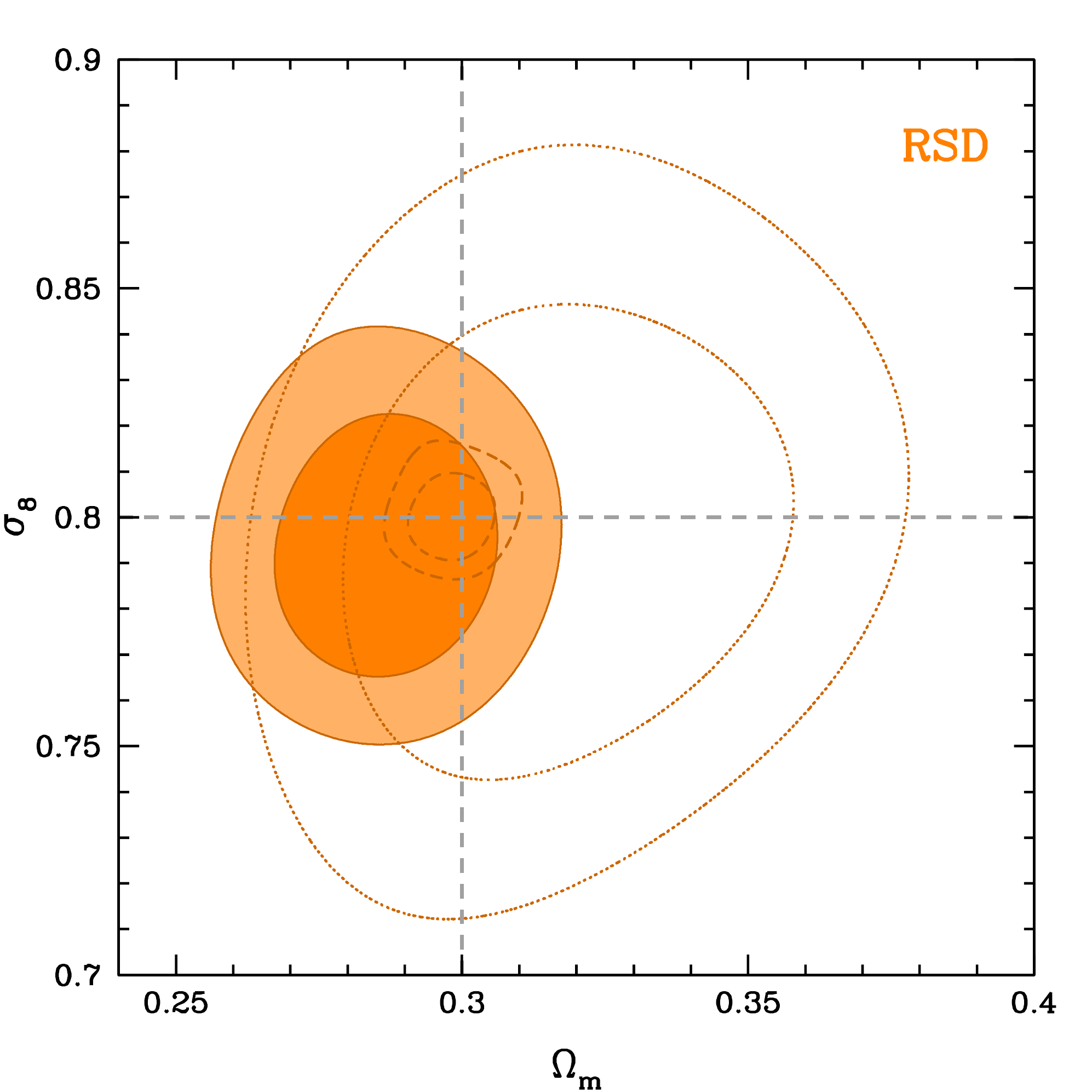}
\includegraphics[scale=0.34]{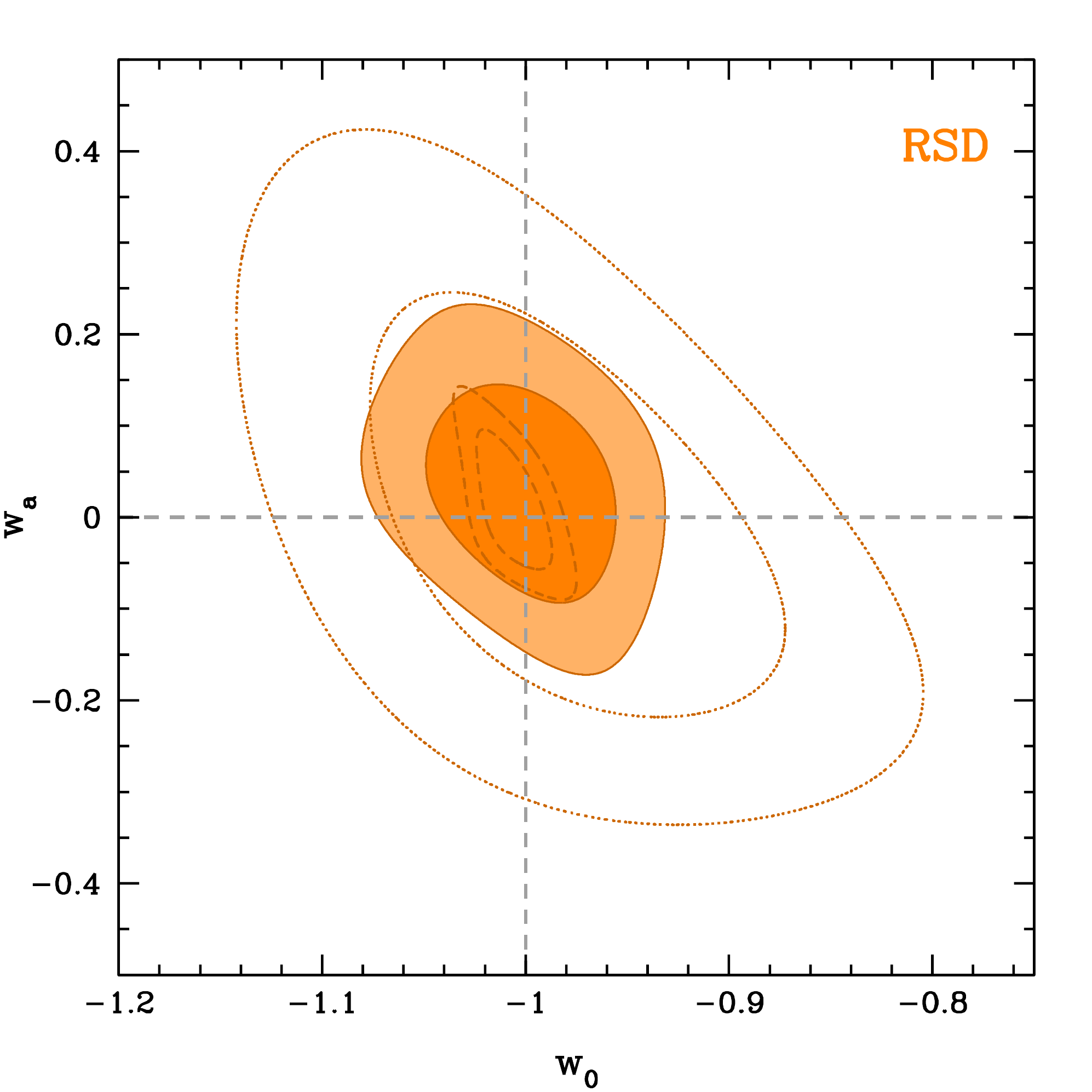}}
\centerline{
\includegraphics[scale=0.34]{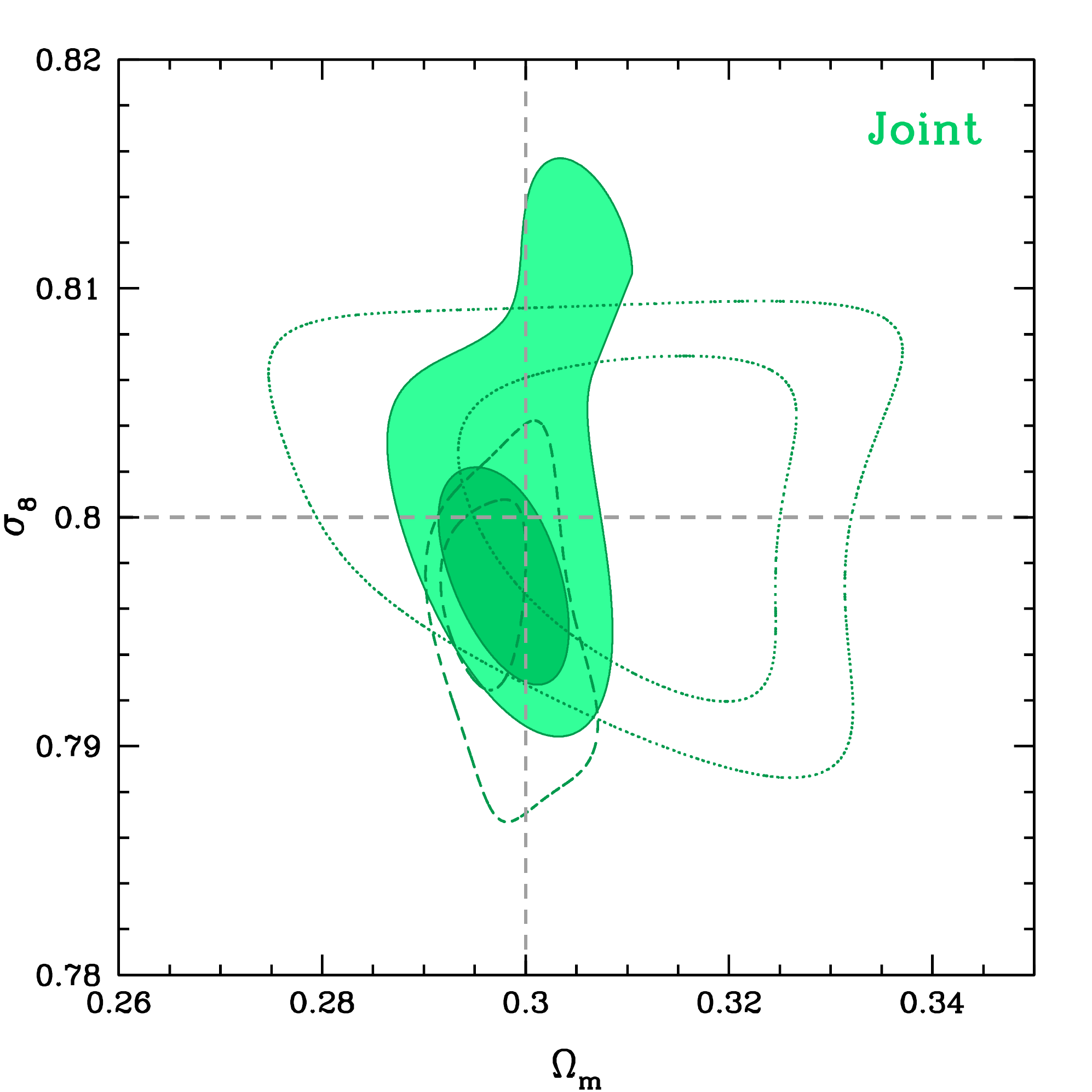}
\includegraphics[scale=0.34]{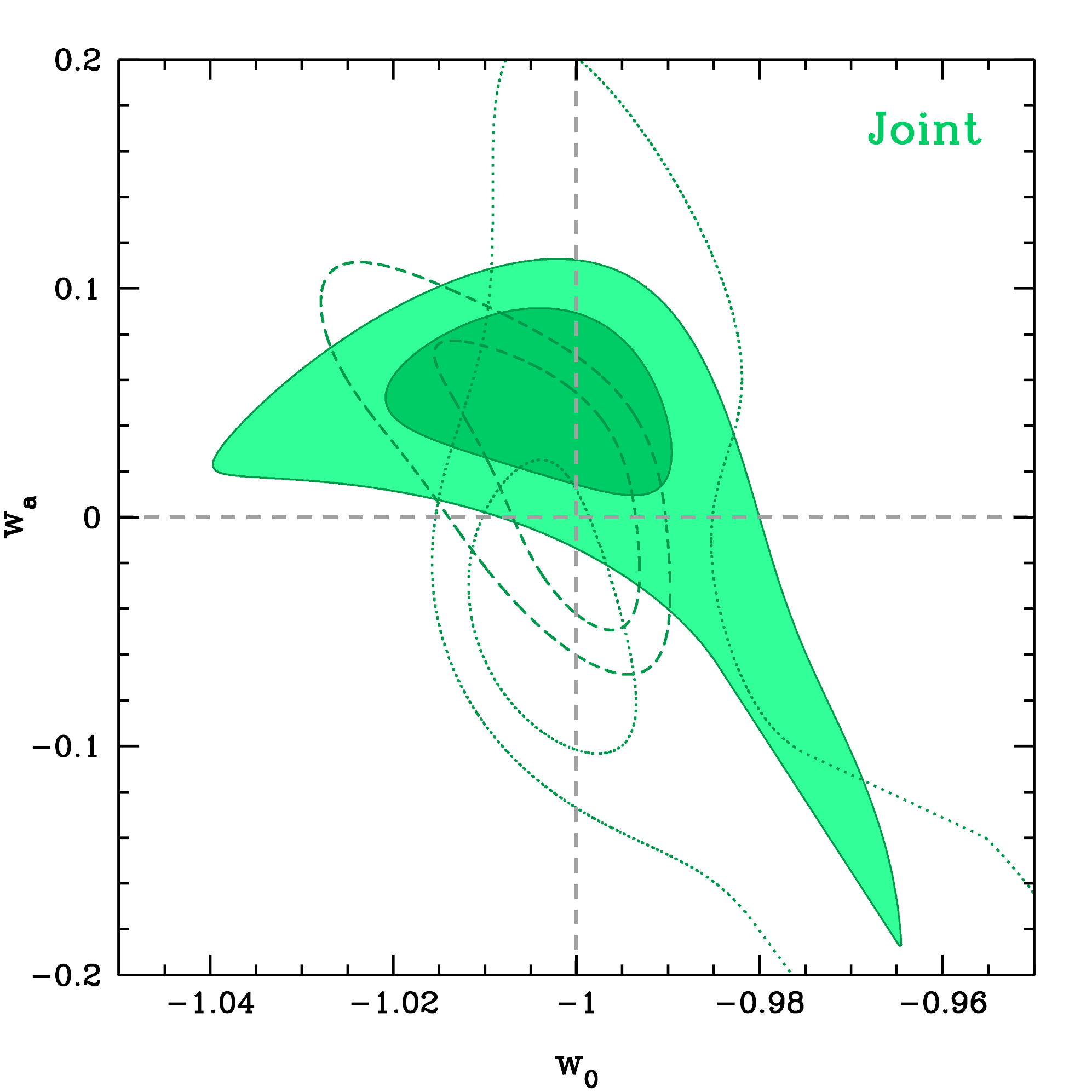}}
\caption{The constraint results of $\Omega_{\rm m}$ vs. $\sigma_8$ and $w_0$ vs. $w_a$ for the CSST WL and galaxy clustering surveys. The $1\sigma$ $(68.3\%)$ and $2\sigma$ $(95.5\%)$ C.L. are shown. $Top:$ results from the WL mock data. The solid, dashed, and dotted contours are the corresponding results by assuming $\bar{N}_{\rm add}=10^{-9}$, $10^{-10}$, and $10^{-8}$, respectively. $Middle:$ the results from the CSST galaxy clustering survey. The solid, dashed, and dotted contours are for $f^{z_{\rm s},0}_{\rm eff}=0.5$, 0.7, and 0.3, and $\bar{N}^g_{\rm sys}=5\times10^4$, $10^4$, and $10^5$ $({\rm Mpc}/h)^3$, respectively. $Bottom:$ Joint constraint results (WL+galaxy clustering+galaxy-galaxy lensing). The three cases (solid, dashed, and dotted) are derived from the moderate, optimistic, and pessimistic assumptions of the WL and galaxy clustering surveys.}
\label{fig:sys_comp}
\end{figure*}

The fiducial value of photo-$z$ bias $\Delta z$ is set to be zero for each photo-$z$ bin, since the outlier fraction in the photo-$z$ fitting is ignored. Note that our method is actually not quite sensitive to the fiducial value of $\Delta z$, that is because we treat it as a free parameter in the fitting process. The fiducial values of $b_g$ for the five bins are obtained by $b_g=b_0(1+z_{\rm c})^{b_1}$, where $b_0=1$ and $b_1=1$ with the central redshifts of the five spec-$z$ bins $z_{\rm c}=0.15$, 0.45, 0.75, 1.05, and 1.35, respectively. The fiducial galaxy velocity dispersion for the five spec-$z$ bins is assumed as $\sigma_v=\sigma_{v,0}/(1+z_{\rm c})$ where $\sigma_{v,0}=7$ ${\rm Mpc}/h$. $s^i_z$ is the stretch factor that adjusts the width of the $n_i(z)$ in a given photo-$z$ bin, which can equally change the redshift variance $\sigma_z$. We set the fiducial values of $\Delta z_i$, ${\rm log_{10}}(s^i_z)$, and $m_i$ to be 0 in the four photometric bins.

The $\chi^2$ statistic method is adopted to fit the mock data, which is defined by
\be
\chi^2 =\sum_{i,j}^{N}\ \left(C_{\rm mock}^{i}-C_{\rm th}^{i}\right) {\rm Cov}_{ij}^{-1} \left(C_{\rm mock}^{j}-C_{\rm th}^{j}\right),
\ee
where $C_{\rm mock}^{i}$ is the mock data, and $C_{\rm th}^{i}$ is the theoretical power spectrum of the $i$th redshift bin. Cov$_{ij}$ is the corresponding covariance matrix. The total $\chi^2$ for joint surveys of the WL and galaxy clustering is given by $\chi^2_{\rm tot}=\chi^2_g + \chi^2_{\gamma} + \chi^2_{g\gamma}$, where $\chi^2_g$, $\chi^2_{\gamma}$, and $\chi^2_{g\gamma}$ are the chi-squares for the galaxy clustering, WL, and galaxy-galaxy lensing power spectra, respectively. The likelihood function then can be calculated by $\mathcal{L}\sim{\rm exp}(-\chi^2/2)$.

\begin{figure}[t]
\includegraphics[scale = 0.42]{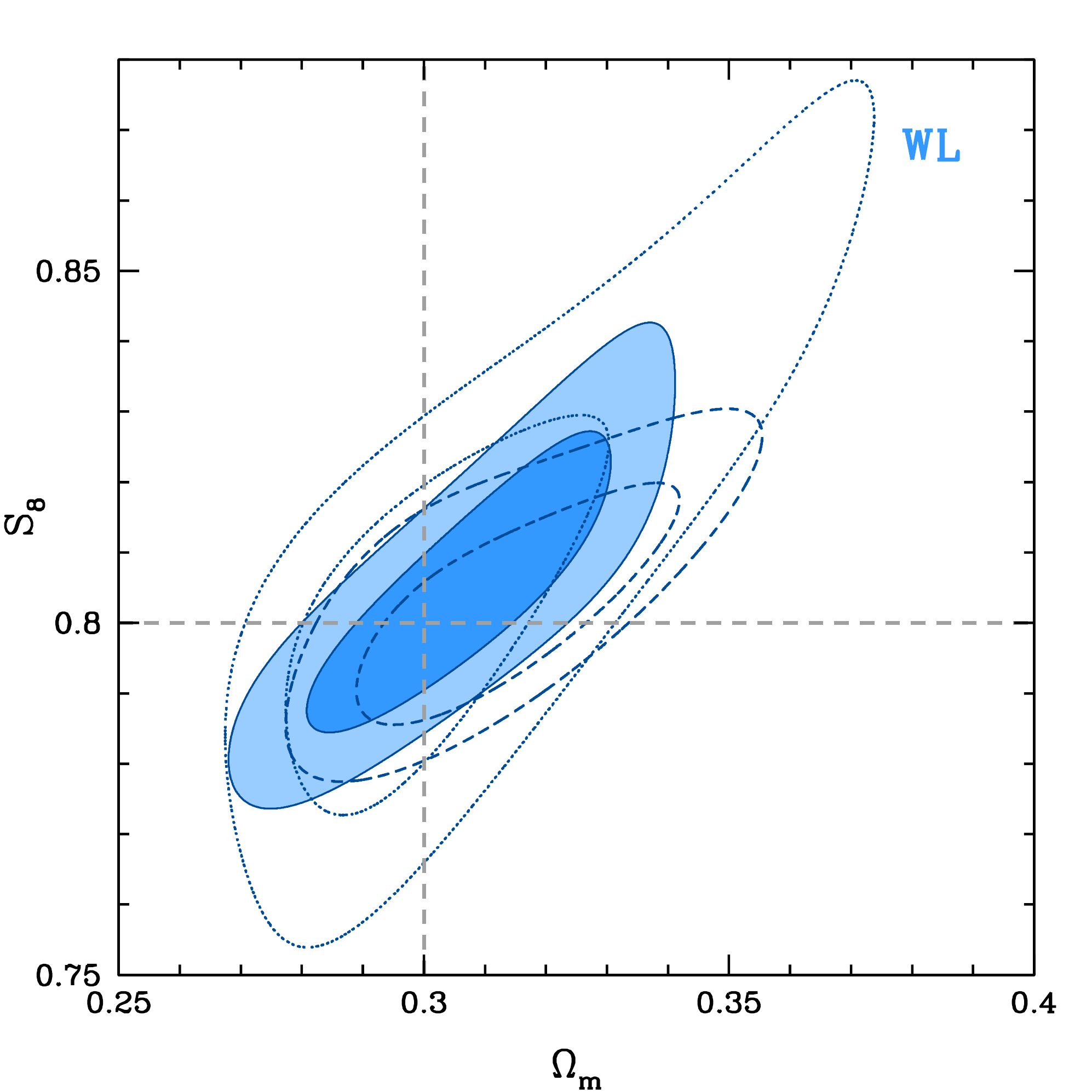}
\caption{\label{fig:m0_SS8} The contour maps ($1\sigma$ and $2\sigma$) of $\Omega_{\rm m}$ vs. $S_8\equiv\sigma_8(\Omega_m/0.3)^{0.5}$ for $\bar{N}_{\rm add}=10^{-9}$ (solid), $10^{-10}$ (dashed), and $10^{-8}$ (dotted), respectively.}
\end{figure}

We make use of the Markov Chain Monte Carlo (MCMC) technique to constrain the free parameters in the model.  The Metropolis-Hastings algorithm is adopted to find the accepting probability of new chain points \citep{Metropolis53,Hastings70}. The proposal density matrix is estimated by a Gaussian sampler with adaptive step size \citep{Doran04}. We assume flat priors for the parameters as shown in Table~\ref{tab:parameters}. We run sixteen parallel chains for each case of systematical assumption, and obtain about 100,000 points for one chain after reaching the convergence criterion \citep{Gelman92}. After the burn-in and thinning processes, we combine all chains together and obtain about 10,000 chain points to illustrate one-dimensional (1-d) and 2-d probability distribution functions (PDFs) of the free parameters.

\begin{figure*}[t]
\centerline{
\resizebox{!}{!}{\includegraphics[scale=0.41]{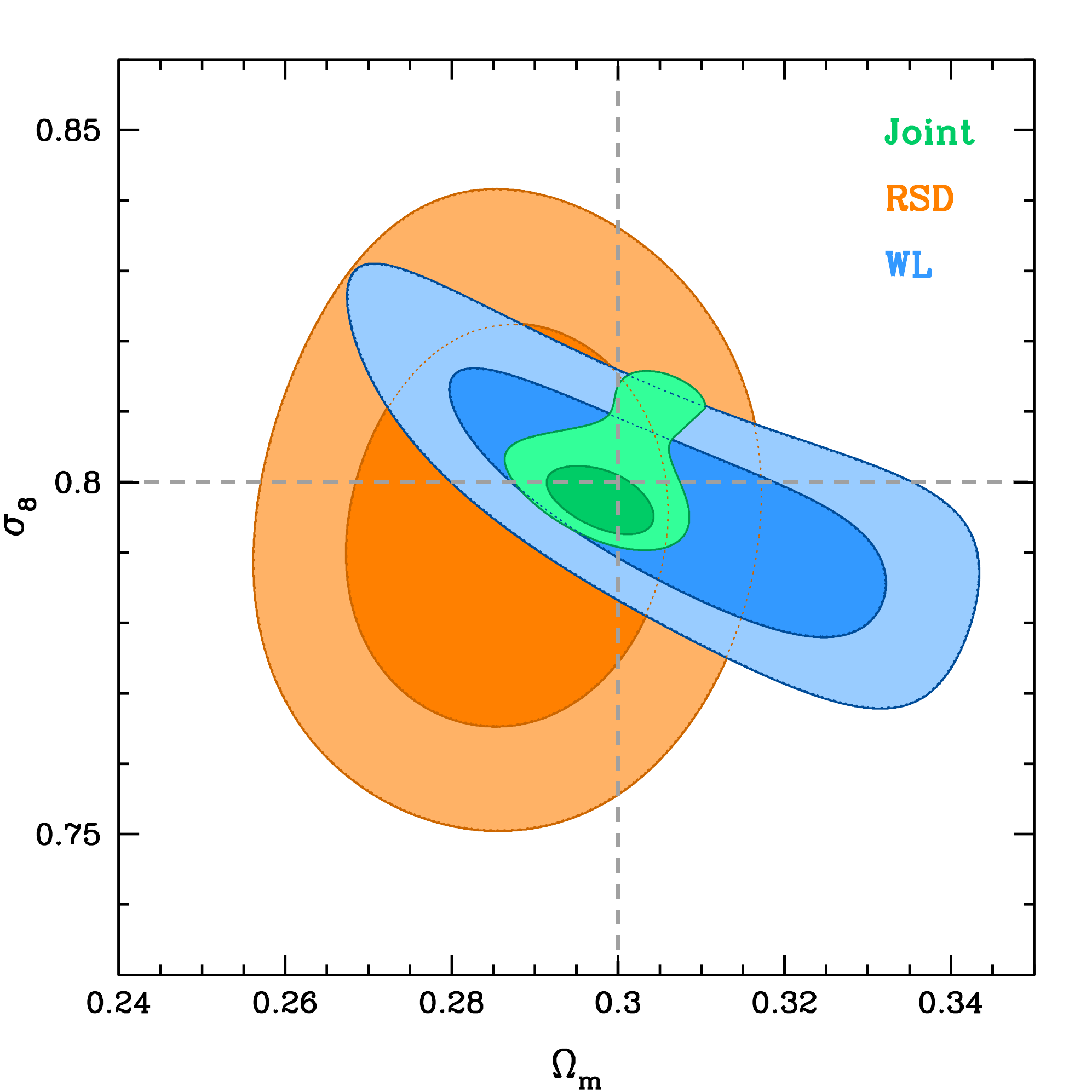}}
\resizebox{!}{!}{\includegraphics[scale=0.41]{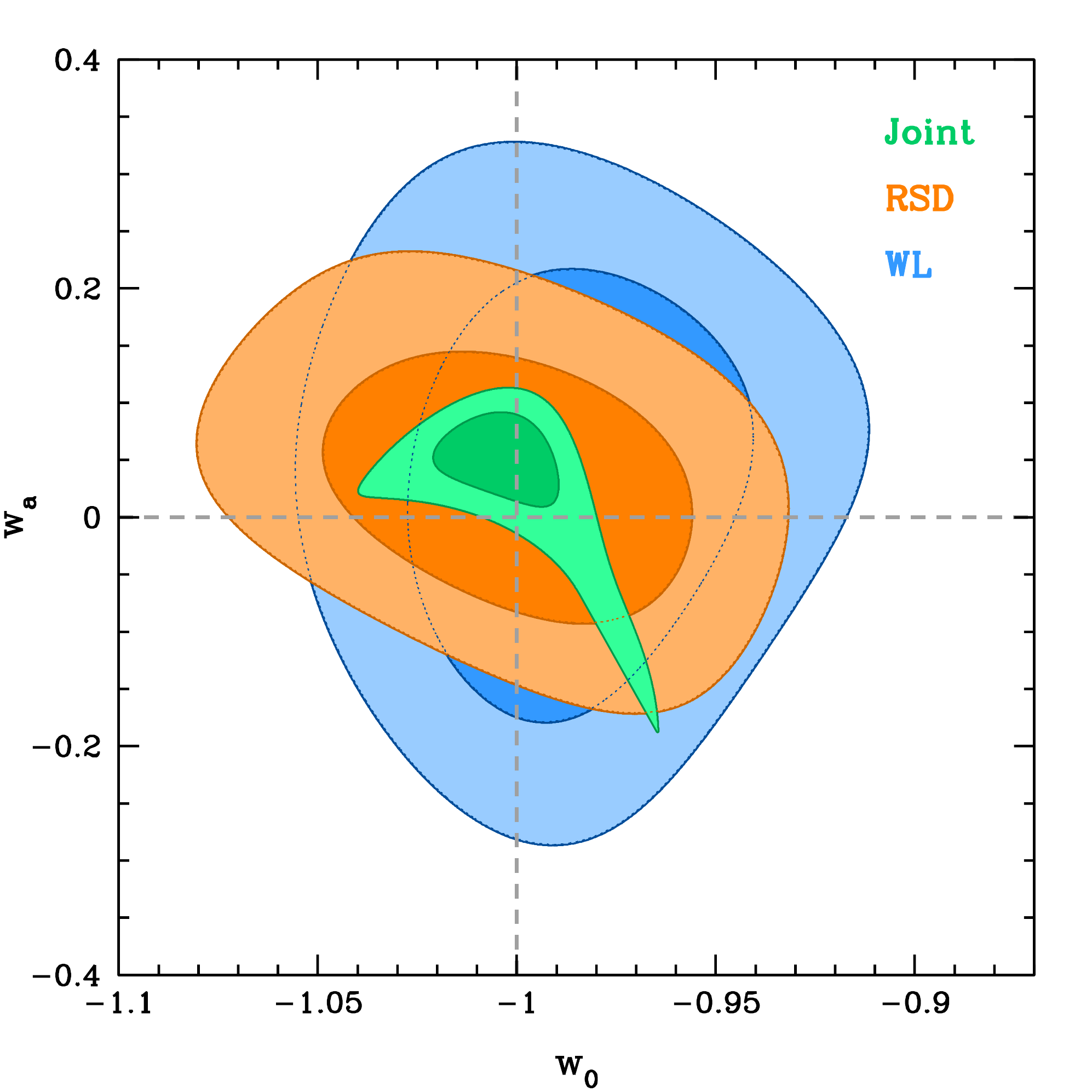}}
}
\caption{Comparison of the constraint results of $\Omega_{\rm m}$ vs. $\sigma_8$ (left panel) and $w_0$ vs. $w_a$ (right panel) for the WL, galaxy clustering, and joint surveys. The $1\sigma$ $(68.3\%)$ and $2\sigma$ $(95.5\%)$ C.L. are shown. Here we assume the moderate systematics, i.e. $\bar{N}_{\rm add}=10^{-9}$, $f^{z_{\rm s},0}_{\rm eff}=0.5$, and $\bar{N}^g_{\rm sys}=5\times10^4$ $({\rm Mpc}/h)^3$.}
\label{fig:data_comp}
\end{figure*}

\section{Constraint results}

In this section, we show the constraint results of the cosmological and systematical parameters using the CSS-OS mock data. We compare the results in different cases for the cosmological parameters, and discuss the impact of the systematics on the constraint results.

\subsection{Constraints on cosmological parameters}

In Figure~\ref{fig:sys_comp}, the fitting results of $\Omega_{\rm m}$ vs. $\sigma_8$ and $w_0$ vs. $w_a$ have been shown. We explore three cases, i.e. the moderate, optimistic, and pessimistic assumptions (in solid, dashed, and dotted curves), about the systematics of the CSST WL and galaxy clustering surveys. The top, middle, and bottom panels show the constraint results from the WL, galaxy clustering, and joint  (WL+galaxy clustering+galaxy-galaxy lensing) surveys, respectively. The gray lines show the fiducial values of the parameters.

For the CSST WL survey, we find that $\Omega_{\rm m}=0.304^{+0.009}_{-0.008}$, $\sigma_8=0.796^{+0.010}_{-0.010}$, $w_0=-1.003^{+0.025}_{-0.021}$, $w_a=-0.013^{+0.078}_{-0.073}$ in $1\sigma$ confidence level (C.L.) for the moderate systematic case with $\bar{N}_{\rm add}=10^{-9}$. We also show the constraint results of $\Omega_{\rm m}$ vs. $S_8\equiv\sigma_8(\Omega_m/0.3)^{0.5}$ in Figure~\ref{fig:m0_SS8}. We find that $S_8=0.801^{+0.009}_{-0.008}$ in the moderate case (solid lines). This constraint result of the cosmological parameters is averagely improved by a factor of $\sim6$ than that from the Dark Energy Survey (DES) and Kilo Degree Survey (KiDS) \citep{Hildebrandt16,Troxel17}. The improvement can be even larger in the optimistic case ($\bar{N}_{\rm add}=10^{-10}$), and at least a factor of $\sim3$ in the pessimistic case ($\bar{N}_{\rm add}=10^{-8}$). This enhancement is due to several advantages of the CSST WL survey, e.g the large survey area (17500 deg$^2$), excellent image quality (small and regular PSF)\footnote{We should note that there are a number of factors can affect the PSF shape in the real observation, coming from the optical system (e.g. optical alignment and thermal stability) and the pointing and steering stability of the telescope. Thus the PSF shape can change in flight, and it could affect the relevant shape measurements in the weak lensing survey and make the systematics larger. Hence, we need to calibrate it to the stated accuracy in the orbit. We will discuss this in details in the future work.}, and accurate photo-$z$ calibration, which can efficiently suppress both statistical and systematical errors.

\begin{figure}
\includegraphics[scale = 0.42]{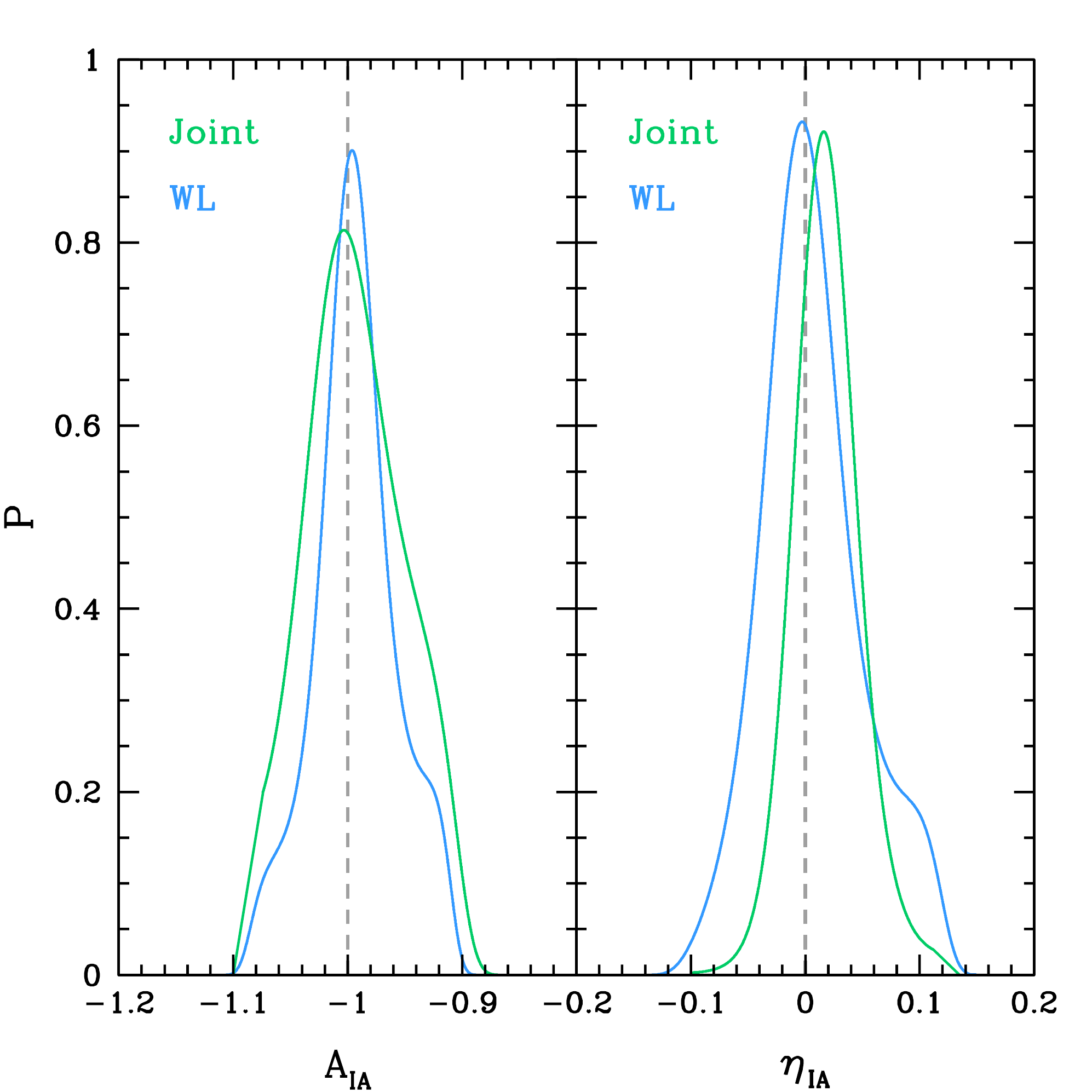}
\caption{The 1-d PDFs of the intrinsic alignment parameters $A_{\rm IA}$ and $\eta_{\rm IA}$ assuming the moderate systematics. The blue and green curves denotes the fitting results of the WL only and joint surveys, respectively.}
\label{fig:A_eta}
\end{figure}

In the CSST galaxy clustering survey, we find that $\Omega_{\rm m}=0.286^{+0.012}_{-0.012}$, $\sigma_8=0.796^{+0.017}_{-0.019}$, $w_0=-1.005^{+0.029}_{-0.031}$, $w_a=0.021^{+0.061}_{-0.050}$ in $1\sigma$ C.L. for the moderate systematic case ($f^{z_{\rm s},0}_{\rm eff}=0.5$ and $\bar{N}^g_{\rm sys}=5\times10^4$ $({\rm Mpc}/h)^3$). This leads to a factor of $\sim5$ improvement at least compared to the current galaxy clustering surveys, such as the SDSS-III Baryon Oscillation Spectroscopic Survey (BOSS) , WiggleZ, 2-degree Field Lensing Survey (2dFLenS), etc. \citep{Zhao16,Wang16,Blake16,Hinton17}. The constraint results could be better or worse with an average factor of 2-3 in the optimistic ($f^{z_{\rm s},0}_{\rm eff}=0.7$ and $\bar{N}^g_{\rm sys}=1\times10^4$ $({\rm Mpc}/h)^3$) and pessimistic ($f^{z_{\rm s},0}_{\rm eff}=0.3$ and $\bar{N}^g_{\rm sys}=1\times10^5$ $({\rm Mpc}/h)^3$) cases. The improvement is mainly caused by that the CSST spectroscopic survey has deep magnitude limit (see Table~\ref{tab:Des_paramts}) and large effective survey volume (see Figure~\ref{fig:Veff}).

\begin{figure*}
\centerline{
\includegraphics[scale=0.27]{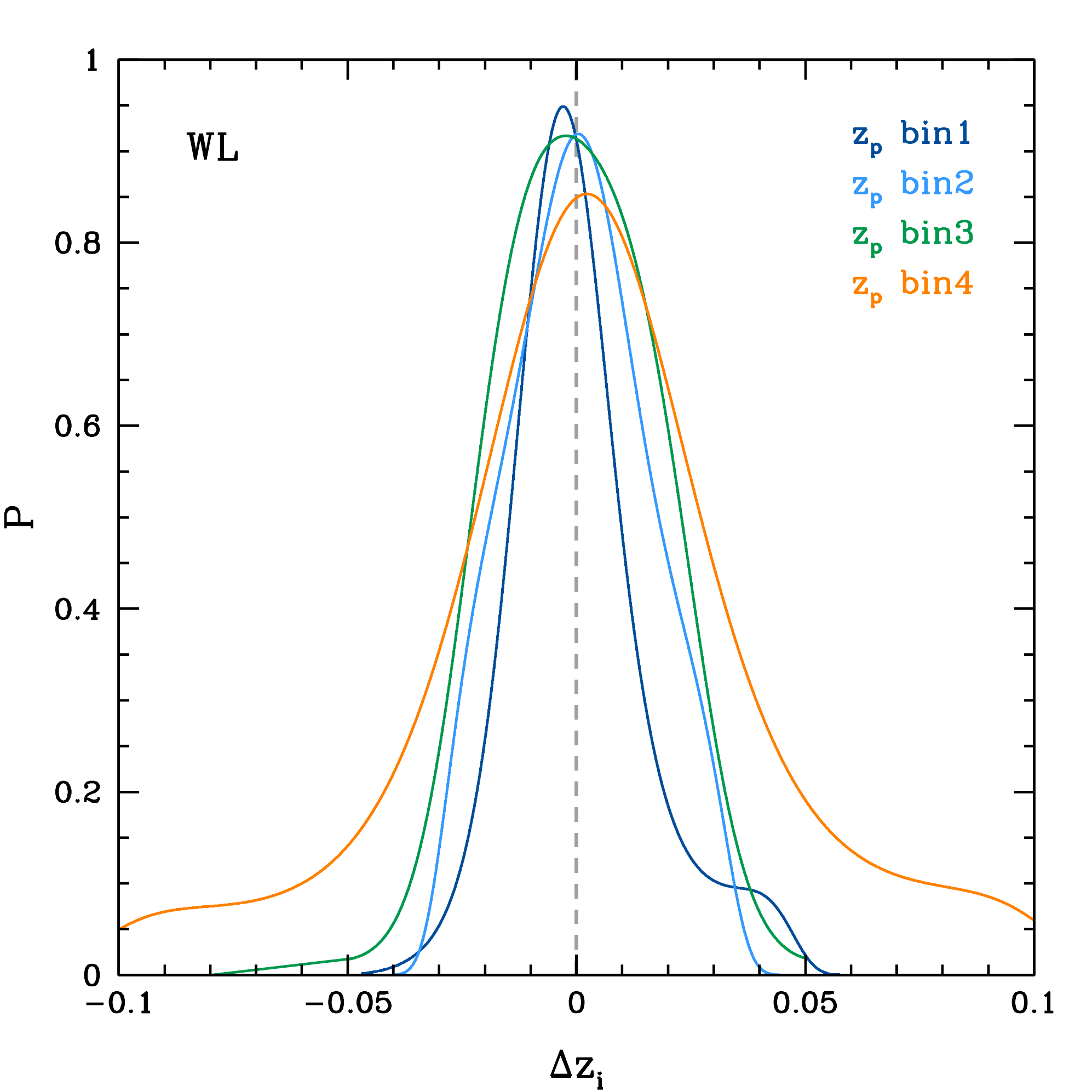}
\includegraphics[scale=0.27]{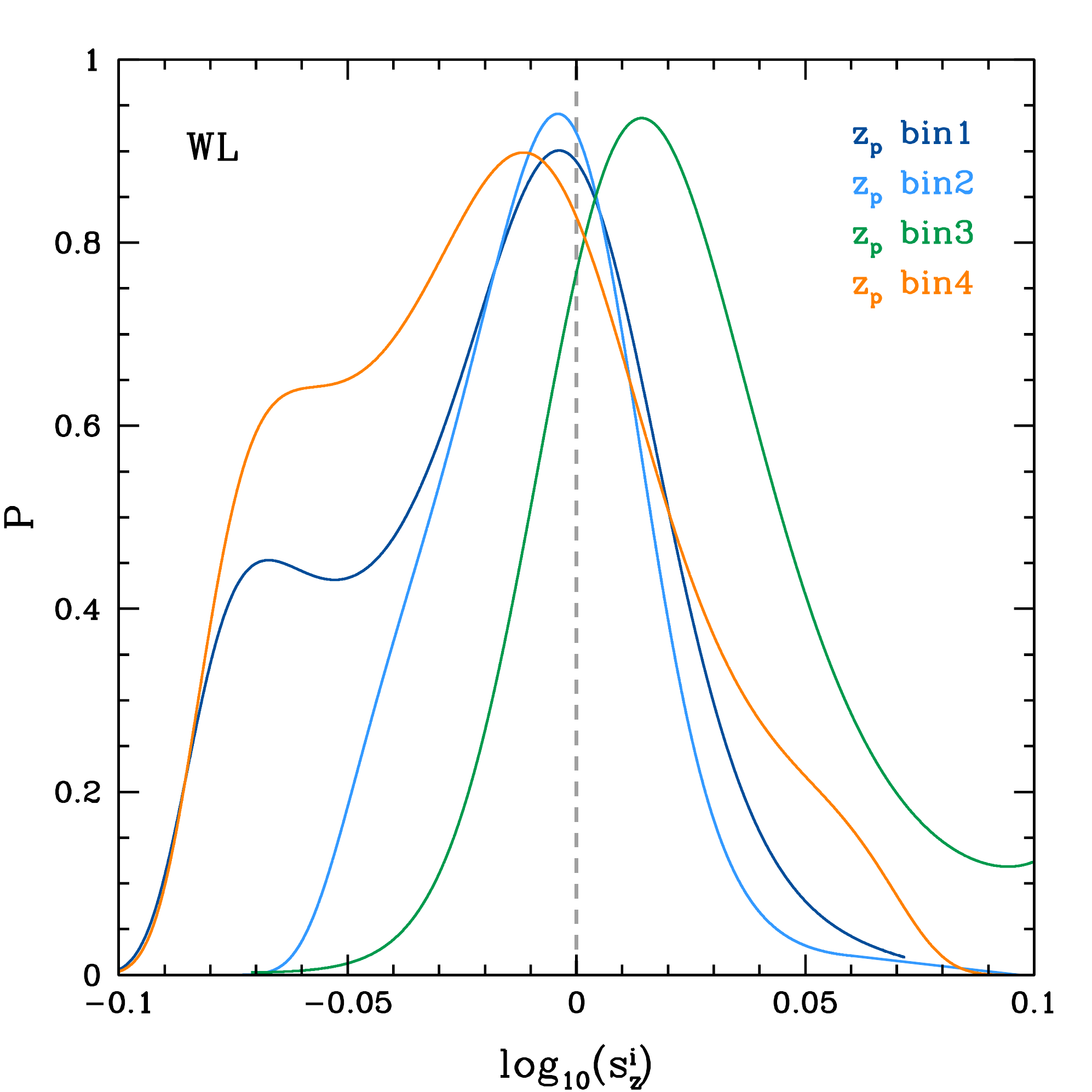}
\includegraphics[scale=0.27]{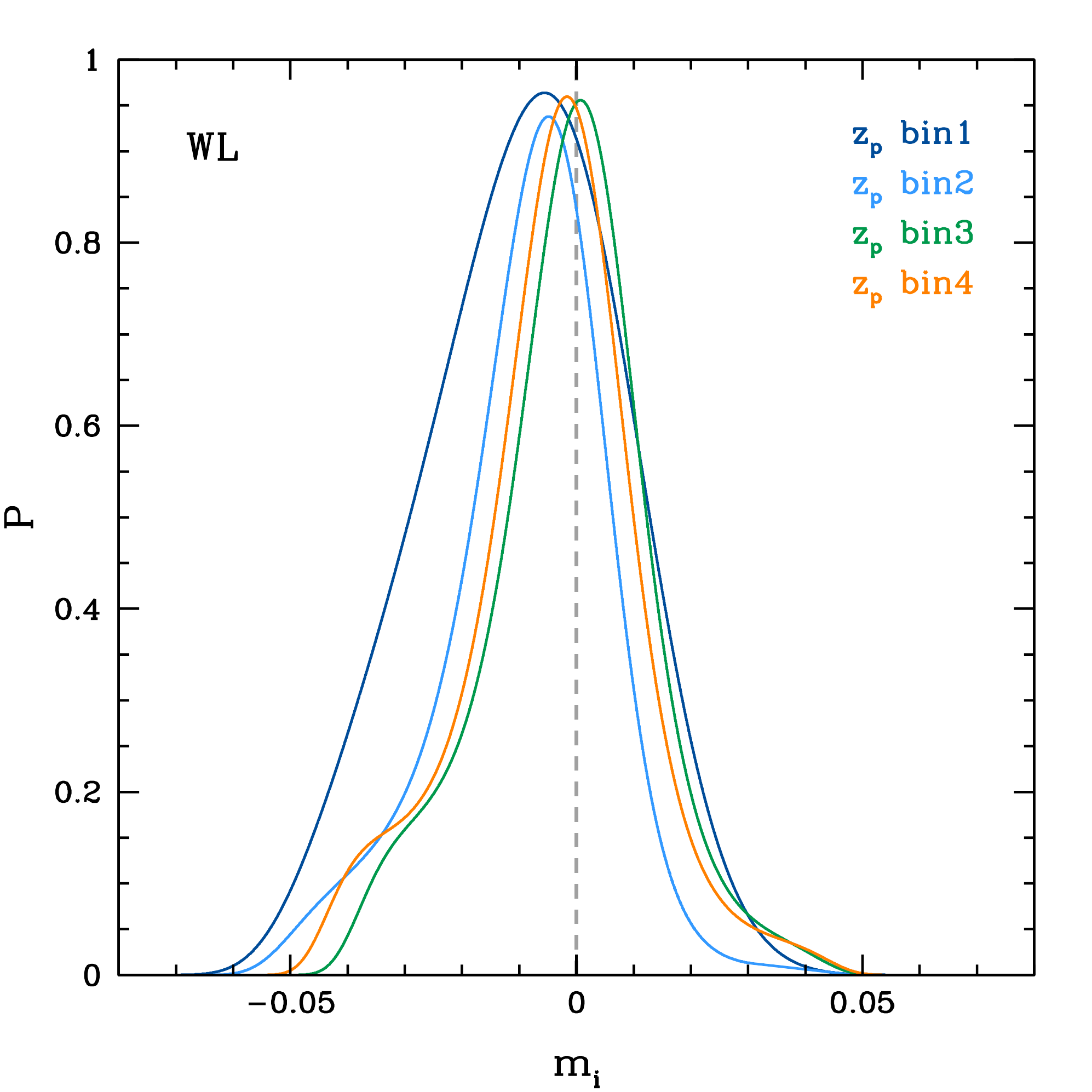}}
\centerline{
\includegraphics[scale=0.27]{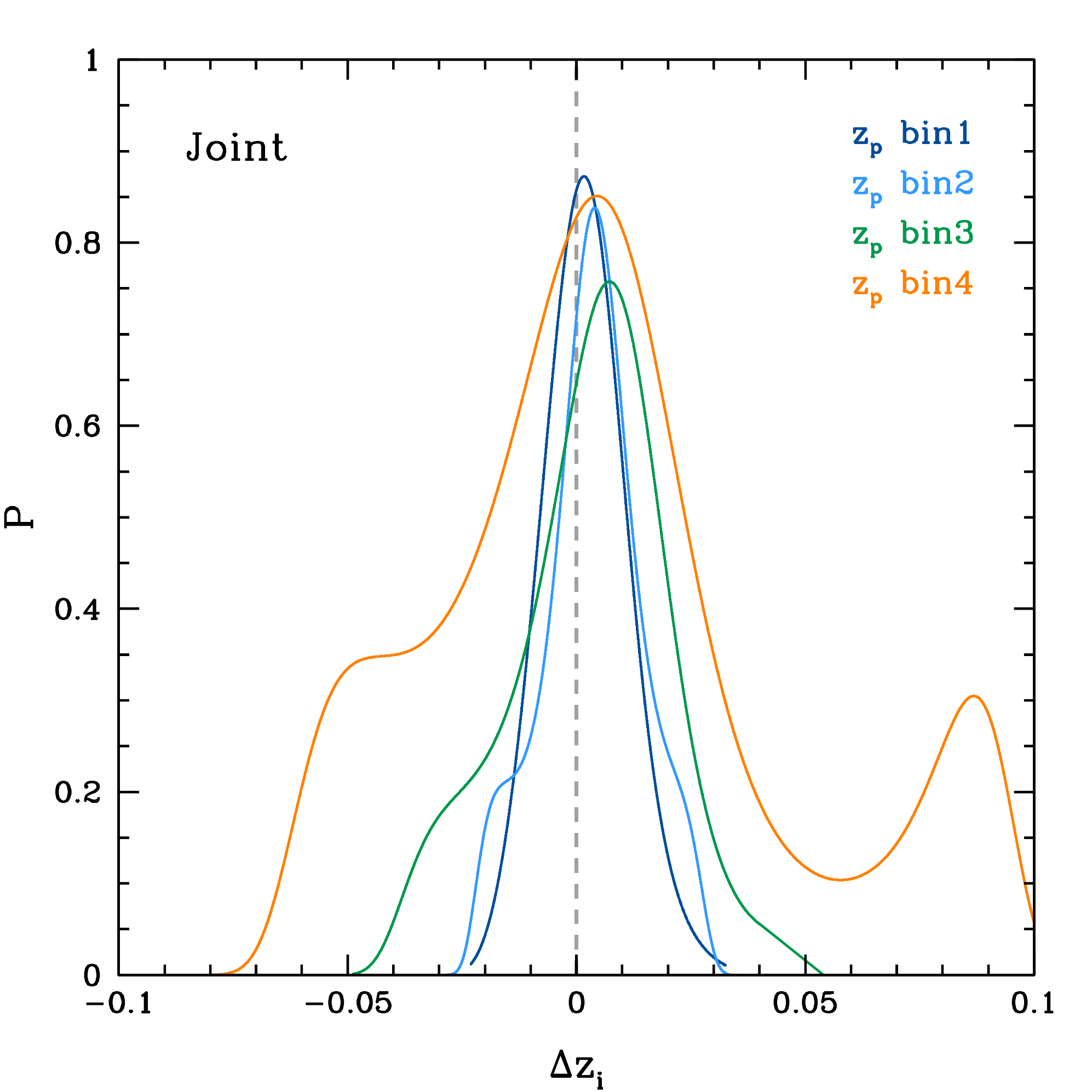}
\includegraphics[scale=0.27]{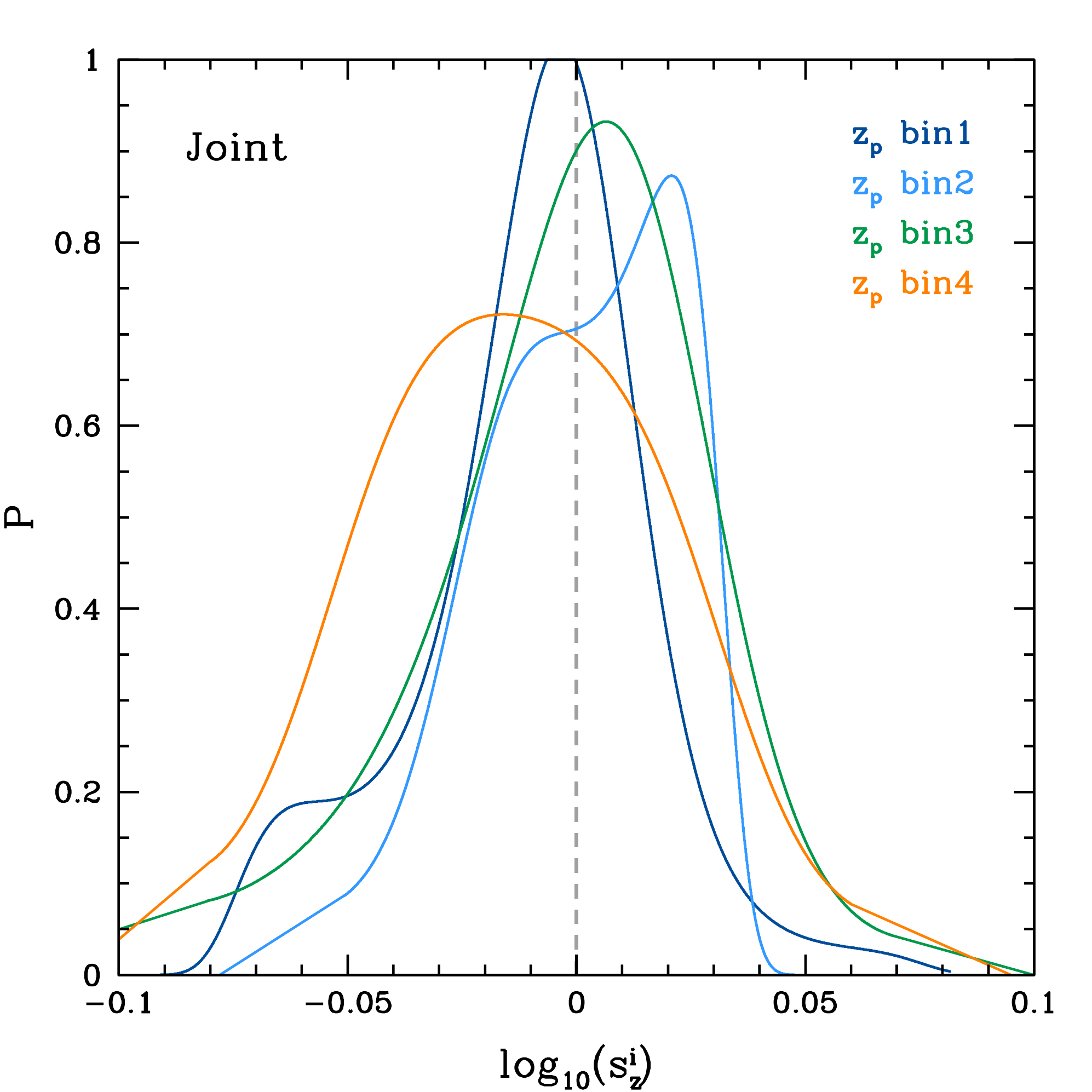}
\includegraphics[scale=0.27]{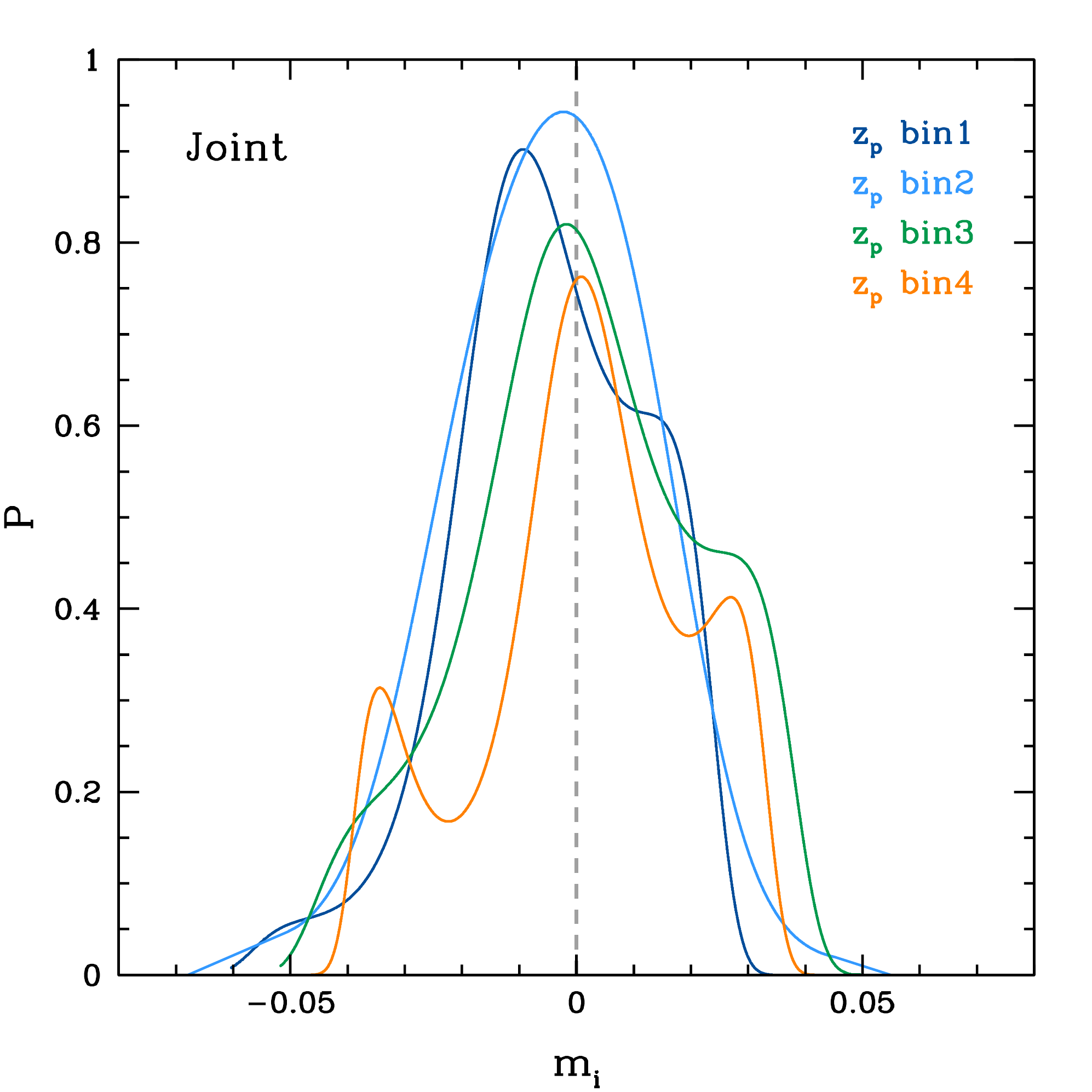}}
\caption{The 1-d PDFs of the systematical parameters $\Delta z_i$, $s_z^i$, and $m_i$ from the CSST WL (top) and joint (bottom) surveys for the four photo-$z$ bins assuming the moderate systematics.}
\label{fig:WL_sys}
\end{figure*}

Since the CSST photometric and spectroscopic surveys are planned to perform simultaneously and cover the same sky area, their joint constraint would be convenient and powerful, which gives $\Omega_{\rm m}=0.304^{+0.003}_{-0.010}$, $\sigma_8=0.803^{+0.005}_{-0.008}$, $w_0=-0.996^{+0.010}_{-0.023}$, $w_a=0.044^{+0.034}_{-0.038}$ in $1\sigma$ C.L. for the moderate systematic case. For the optimistic and pessimistic cases, the constraints can be tighter or looser by a factor of $\sim1.3$ and $1.7$, respectively. Comparing to current similar joint fitting results, e.g. KiDS-450+2dFLenS \citep{Joudaki17}, we find that the CSST joint survey can enhance the constraints of the cosmological parameters by one order of magnitude when assuming the moderate systematics. 

Since the systematics have been included in the analysis, fitting biases with respect to the best-fits appear in the constraint results according to the fiducial values of the cosmological parameters (gray dashed lines), especially for the moderate and pessimistic cases. By comparing the results of the three systematical assumptions, we find that well-controlled systematics not only can shrink the probability contours but also efficiently suppress the fitting biases of the cosmological parameters (see dashed contours in each survey). Besides, it seems that the fitting biases are relatively larger or more apparent in the joint constraint results (e.g. see the result of $w_a$). It means that better controlling of the systematics may be required in the CSS-OS joint fits. The detailed discussion of the systematical parameters can be found in the next section.

In Figure~\ref{fig:data_comp}, we show the comparison of the results from the WL, galaxy clustering, and joint surveys. As can be seen, the CSST WL and galaxy clustering surveys have similar constraint strength on $\Omega_{\rm m}$. On the other hand, the WL is more powerful to constrain $\sigma_8$ than the galaxy clustering by a factor of $\sim$2, since the WL survey explores 2-d power spectra integrating over large redshift range. On the other hand, the CSST galaxy clustering survey can provide comparable or even a bit more stringent fitting results on $w_0$ and $w_a$ than the WL survey. The joint CSST surveys of WL+galaxy clustering+galaxy-galaxy lensing can further improve the fitting results, which give at least $\sim2\sigma$ enhancement on the constraints of cosmological parameters, compared to the WL only or galaxy clustering only survey.

The constraint results of all cosmological parameters with the mild assumption of the systematics for the WL, galaxy clustering, and joint surveys can be found in Appendix.

\subsection{Constraints on systematical parameters}

\begin{figure*}
\centerline{
\includegraphics[scale=0.34]{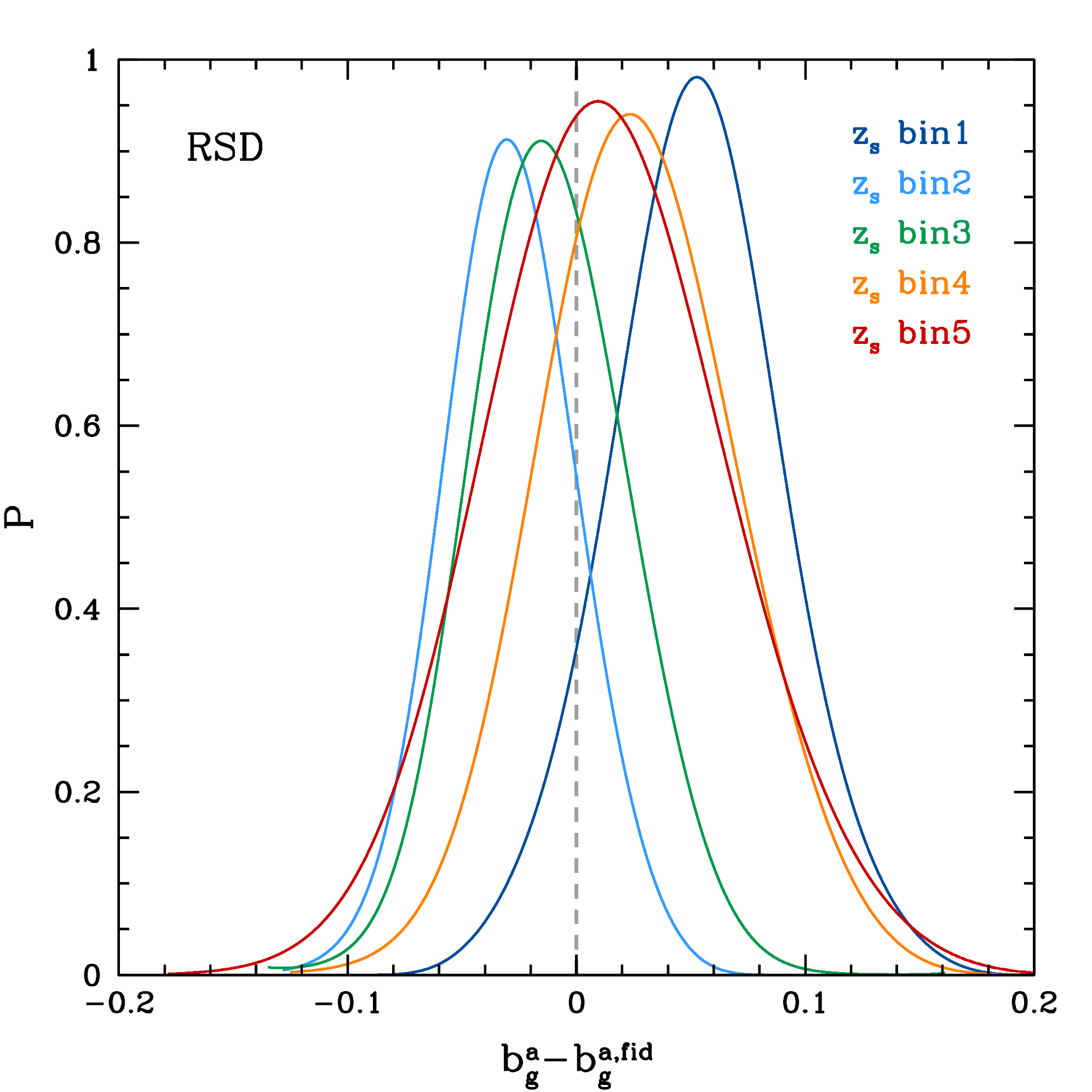}
\includegraphics[scale=0.34]{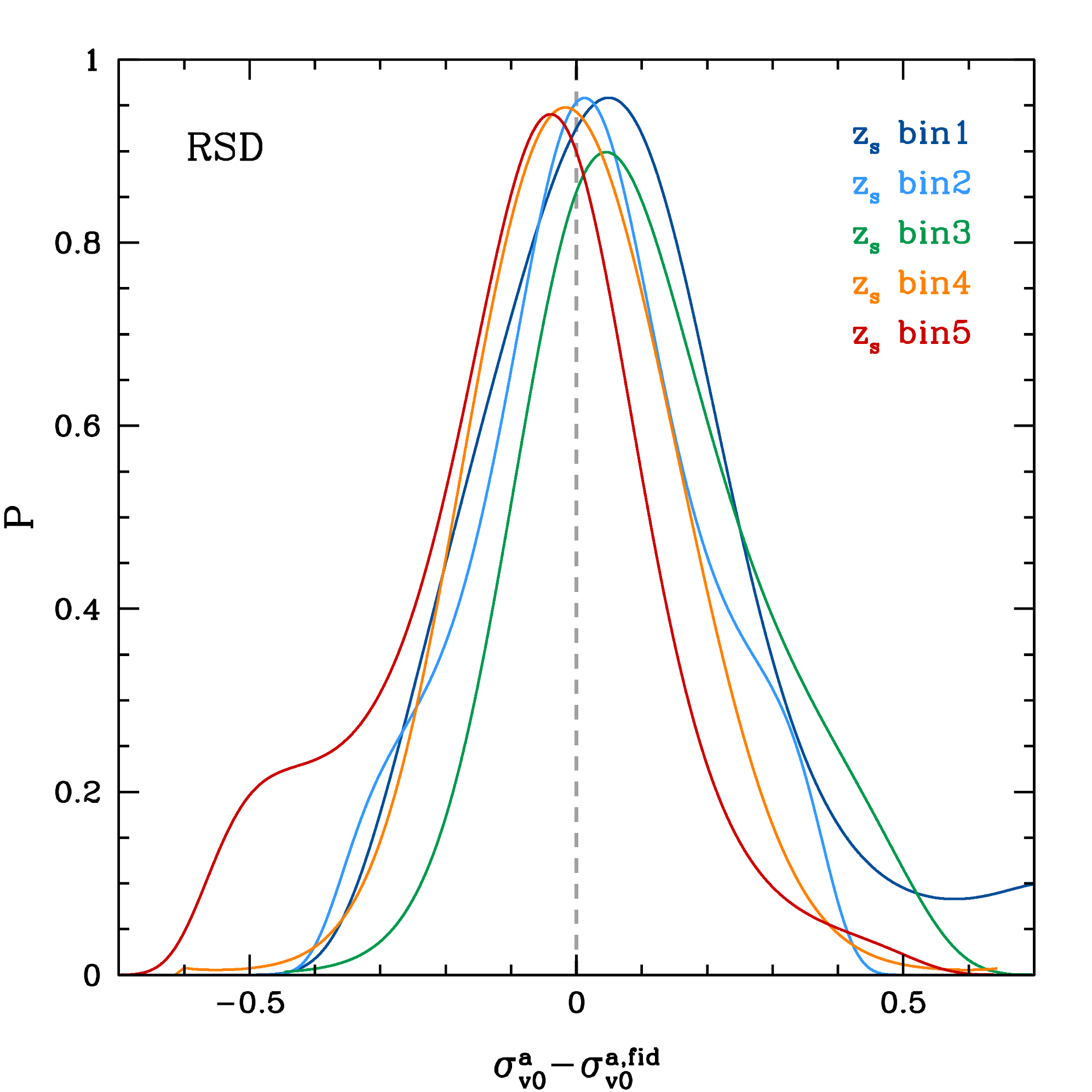}}
\centerline{
\includegraphics[scale=0.34]{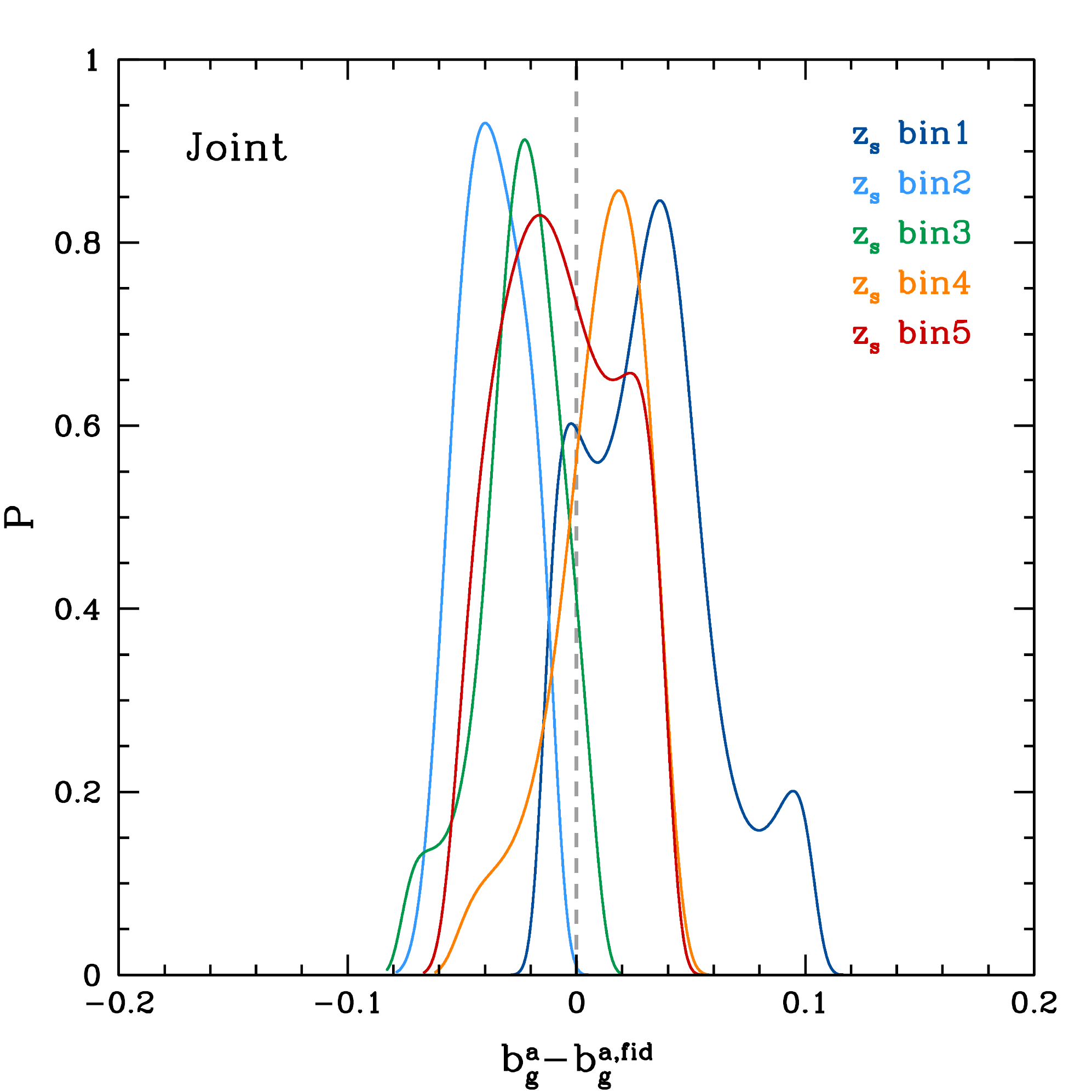}
\includegraphics[scale=0.34]{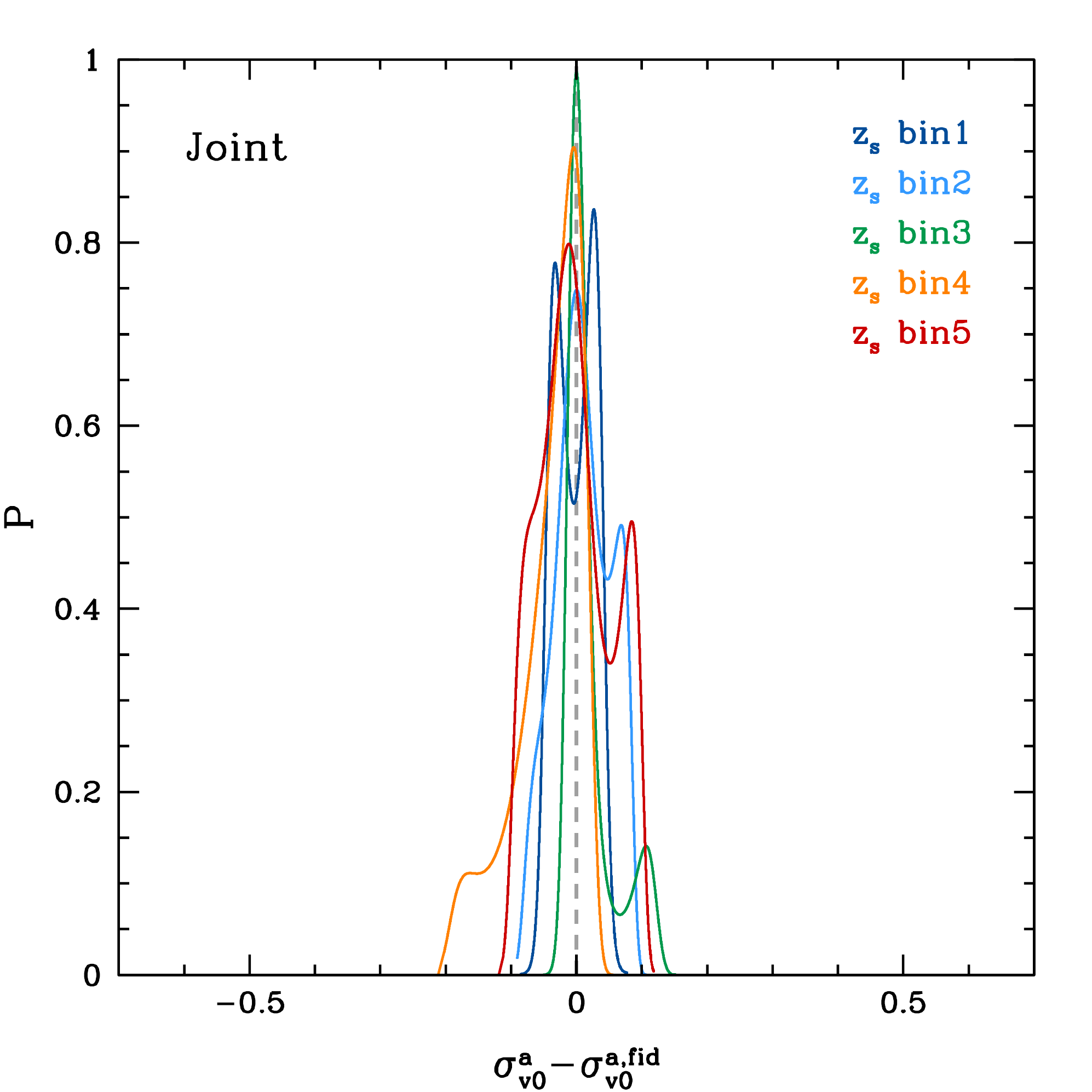}
}
\caption{The 1-d PDFs of galaxy bias $b_g^a$ and velocity dispersion parameter $\sigma_{v0}^a$, compared to their fiducial values, from the CSST galaxy clustering (top) and joint (bottom) surveys for the five spec-$z$ bins assuming the moderate systematics.}
\label{fig:RSD_sys}
\end{figure*}

In Figure~\ref{fig:A_eta}, the 1-d PDFs of the intrinsic alignment parameters $A_{\rm IA}$ and $\eta_{\rm IA}$ are shown for the WL (in blue curves) and joint (in green curves) surveys. We can find that there is no significant improvement on the constraint of $A_{\rm IA}$ for the joint constraints, while a factor of $\sim1.3$ tighter for $\eta_{\rm IA}$ compared to the WL survey. This indicates that the joint fitting can be helpful to extract the redshift evolution effect of intrinsic alignment.

The 1-d PDFs of the redshift calibration bias $\Delta z_i$, stretch factor $s_z^i$, and multiplicative error $m_i$, are shown in Figure~\ref{fig:WL_sys}. The top and bottom panels show the results from the WL and joint surveys, respectively. We can see that the best-fits of the WL systematic parameters are close to their fiducial values (in 1$\sigma$ C.L.), that means our fitting process can correctly extract the systematics as free parameters. This is also useful to reduce the effect of the systematics on the constraints of cosmological parameters, and can help suppressing the fitting biases (see Figure~\ref{fig:sys_comp}). In order to retain small fitting biases of the cosmological parameters (keeping the fiducial values in 1$\sigma$ C.L.), as shown in the top panels of Figure~\ref{fig:WL_sys}, we need to control the systematical parameters in 1$\sigma$ C.L. of their PDFs at least. This requires $|\Delta z_i|<0.02$, $|{\rm log_{10}}(s_z^i)|<0.05$ (about 10\%)\footnote{The stretch factor of the redshift distribution has less effect compared to the redshift bias, which has not been considered in the current WL data analysis \citep[e.g. see][]{Hildebrandt16,Troxel17}.}, and $|m_i|<0.02$ for the CSST WL survey. When performing the joint fitting, the constraint results of the systematics can be improved by a factor of $\sim1.5$, which implies that it is helpful to include other surveys (here the galaxy clustering survey) to eliminate the WL systematics in the fitting process.

In Figure~\ref{fig:RSD_sys}, the 1-d PDFs of galaxy bias $b_g$ and velocity dispersion $\sigma_{v0}$ assuming the moderate systematics for the five spec-$z$ bins are shown. For the galaxy clustering survey (top panels), we find that the dispersions of the best-fits of $b_g$ and $\sigma_{v0}$ from the fiducial values are within $\pm$0.1, and the $1\sigma$ ranges 0.03-0.05 for $b_g$ and 0.1-0.2 for $\sigma_{v0}$. The dispersions and uncertainties of $b_g$ and $\sigma_{v0}$ can be further suppressed in the joint fitting (bottom panel), especially for $\sigma_{v0}$. We find that the dispersions of $b_g$ and $\sigma_{v0}$ can be restricted within $\pm$0.05, and the $1\sigma$ within 0.01-0.02 for $b_g$ and 0.02-0.06 for $\sigma_{v0}$ (a factor of $\sim$2 and $\sim4$ improvement, respectively.). It indicates that the CSS-OS can provide accurate constraints on the galaxy bias and velocity dispersion.

\section{summary and discussion}

In this work, we predict the measurements of the CSS-OS on the weak gravitational lensing and galaxy clustering, and explore the constraints on the cosmological parameters with the systematics. We make use of two catalogs, i.e. COSMOS and zCOSMOS catalogs, to simulate the CSST photometric imaging and slitless spectroscopic surveys. We find that the peaks of galaxy redshift distributions are around 0.6 and 0.3 for the CSST photometric and spectroscopic surveys, respectively.

We divide the photometric redshift distribution into four bins, and calculate the auto and cross convergence power spectra of these photo-$z$ bins. The effect of intrinsic alignment, and multiplicative and additive errors are included when generating the mock WL data. In addition to the photometric WL survey, the CSST can simultaneously perform spectroscopic survey to illustrate clustering of galaxies. We compute the galaxy redshift-space power spectra in five spec-$z$ bins, and obtain the mock data of the multipole power spectra, i.e. $P_0^g$, $P_2^g$, and $P_4^g$. We consider a number of effects when estimating the errors, such as the frequency resolution of the slitless grating, the effective redshift factor accounting for the fraction of galaxies that can achieve the required redshift accuracy in the slitless observations, and the systematic error due to the instrument effect. The cross correlations of the CSST WL and galaxy clustering surveys, i.e. galaxy-galaxy lensing power spectra, are also explored, which can be helpful to further suppress the uncertainties in the joint surveys.

After obtaining the CSS-OS mock data, the MCMC technique is adopted to constrain the cosmological and systematical parameters. We study and compare three cases with different assumptions about the systematics of the WL and galaxy clustering surveys, i.e. the pessimistic, moderate, and optimistic cases, in the constraint process. We find that the CSST WL and galaxy clustering surveys can provide a factor of a few (and optimistically even one order of magnitude) improvement about the cosmological parameters, compared to the current corresponding surveys. The constraints can be further enhanced by $\sim2\sigma$ in the joint fitting process (WL+galaxy clustering+galaxy-galaxy lensing). We should note that the constraints can be worse in the real survey, since some assumptions made in this work may be simple and optimistic, and more details and uncertainties need to be further considered in future realistic simulations to derive more realistic result.

The CSS-OS also could provide good fitting about the intrinsic alignment and systematics (in the redshift and shape calibrations) in the WL survey, and galaxy bias and velocity dispersion in the galaxy clustering survey. The joint constraint can further improve the results by a factor of $\sim$2-4. Particularly, the systematics should be well controlled to avoid large fitting bias on the cosmological parameters in the WL survey, which requires redshift bias $|\Delta z_i|<0.02$, the redshift stretching scale (or the uncertainty of the redshift variance) $s_z^i<10\%$, and the multiplicative error $|m_i|<0.02$. 

Besides the WL and redshift-space 3-d galaxy clustering surveys discussed above, the CSS-OS also can perform 2-d angular galaxy clustering photometric survey, strong gravitational lensing survey, galaxy cluster survey, etc. These surveys can offer more valuable information about dark matter and dark energy, the evolution of the LSS, and other important issues in cosmology. Therefore, we can expect that the CSS-OS will be a powerful space sky survey for the studies of our Universe.

\begin{acknowledgments}
YG acknowledges the support of NSFC-11822305, NSFC-11773031, NSFC-11633004, the Chinese Academy of Sciences (CAS) Strategic Priority Research Program XDA15020200, the NSFC-ISF joint research program No. 11761141012, and CAS Interdisciplinary Innovation Team. XKL acknowledges the support from YNU Grant KC1710708 and NSFC-11803028. ZHF acknowledges the support of NSFC-11333001. XZ and HZ were partially supported by China Manned Space Program through its Space Application System and by the National Key R\&D Program of China grant No. 2016YFB1000605.
\end{acknowledgments}

\appendix

\begin{figure*}
\centerline{
\resizebox{!}{!}{\includegraphics[scale=0.8]{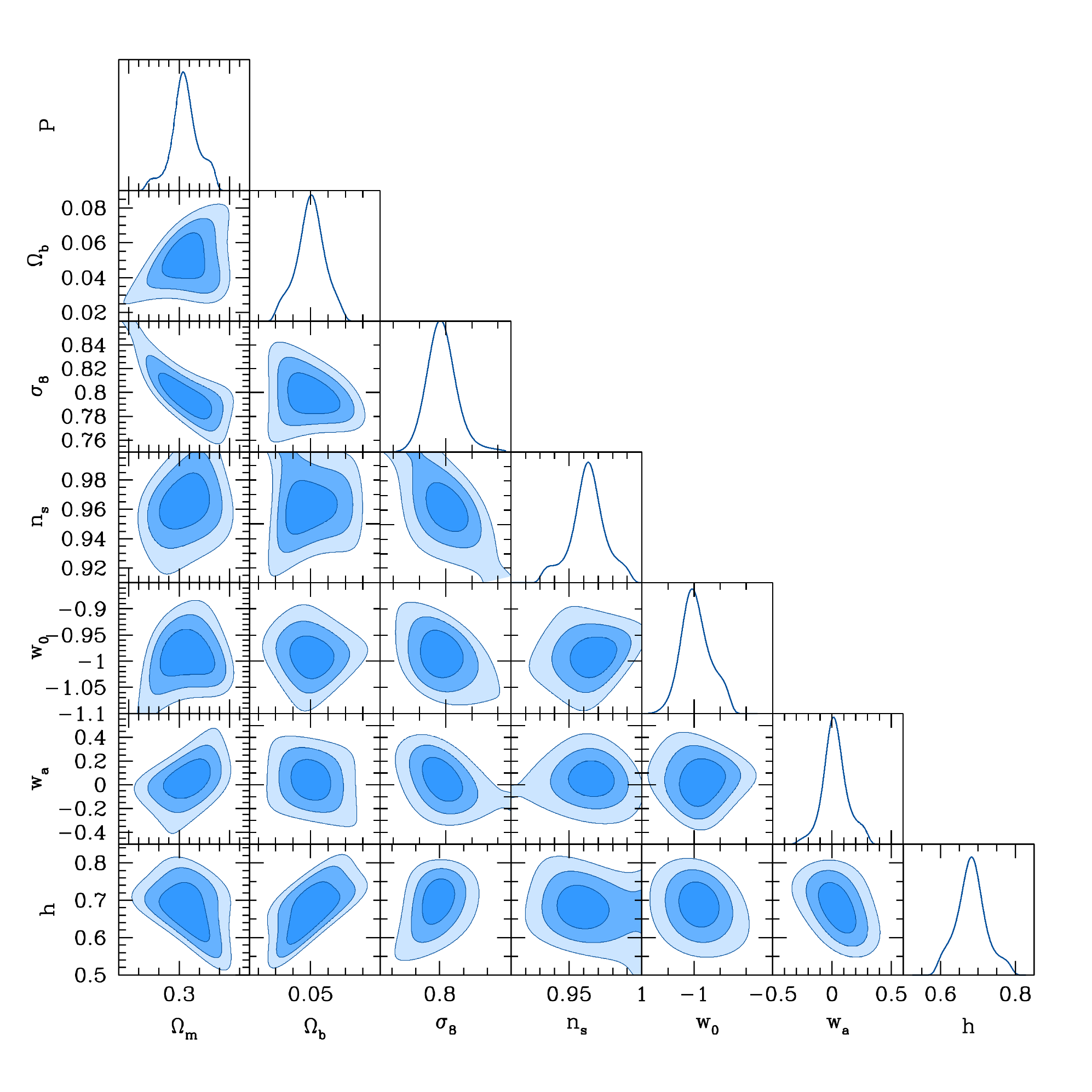}}
}
\caption{The constraint results of the seven cosmological parameters from the CSST WL survey, assuming $\bar{N}_{\rm add}=10^{-9}$. The $1\sigma$ $(68.3\%)$, $2\sigma$ $(95.5\%)$, and $3\sigma$ $(99.7\%)$ C.L. are shown.}
\label{fig:WL_all}
\end{figure*}

\begin{figure*}
\centerline{
\resizebox{!}{!}{\includegraphics[scale=0.8]{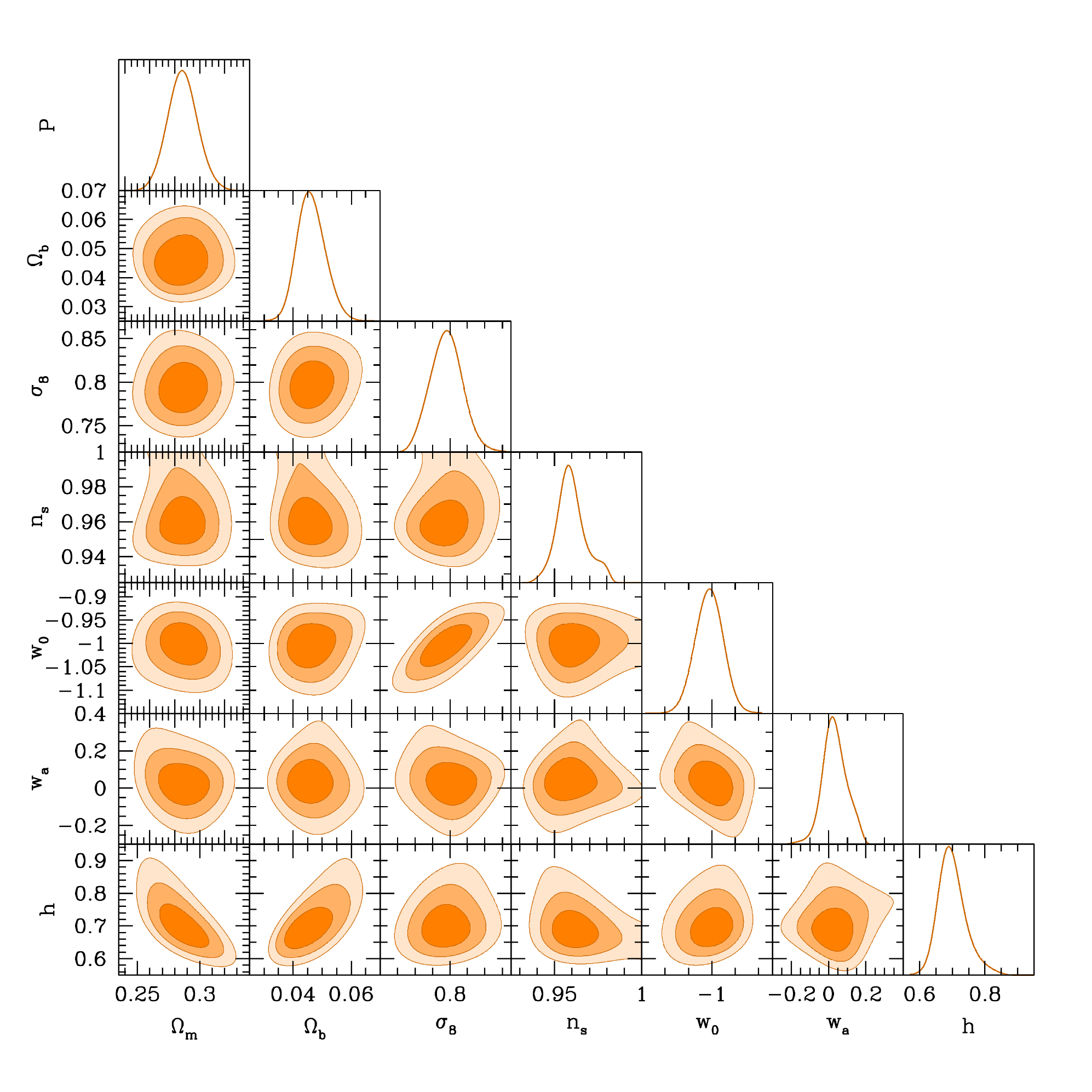}}
}
\caption{The constraint results of the seven cosmological parameters from the CSST galaxy clustering survey, assuming $f^{z_{\rm s},0}_{\rm eff}=0.5$ and $\bar{N}^g_{\rm sys}=5\times10^4$ $({\rm Mpc}/h)^3$. The $1\sigma$ $(68.3\%)$, $2\sigma$ $(95.5\%)$, and $3\sigma$ $(99.7\%)$ C.L. are shown.}
\label{fig:RSD_all}
\end{figure*}

\begin{figure*}
\centerline{
\resizebox{!}{!}{\includegraphics[scale=0.8]{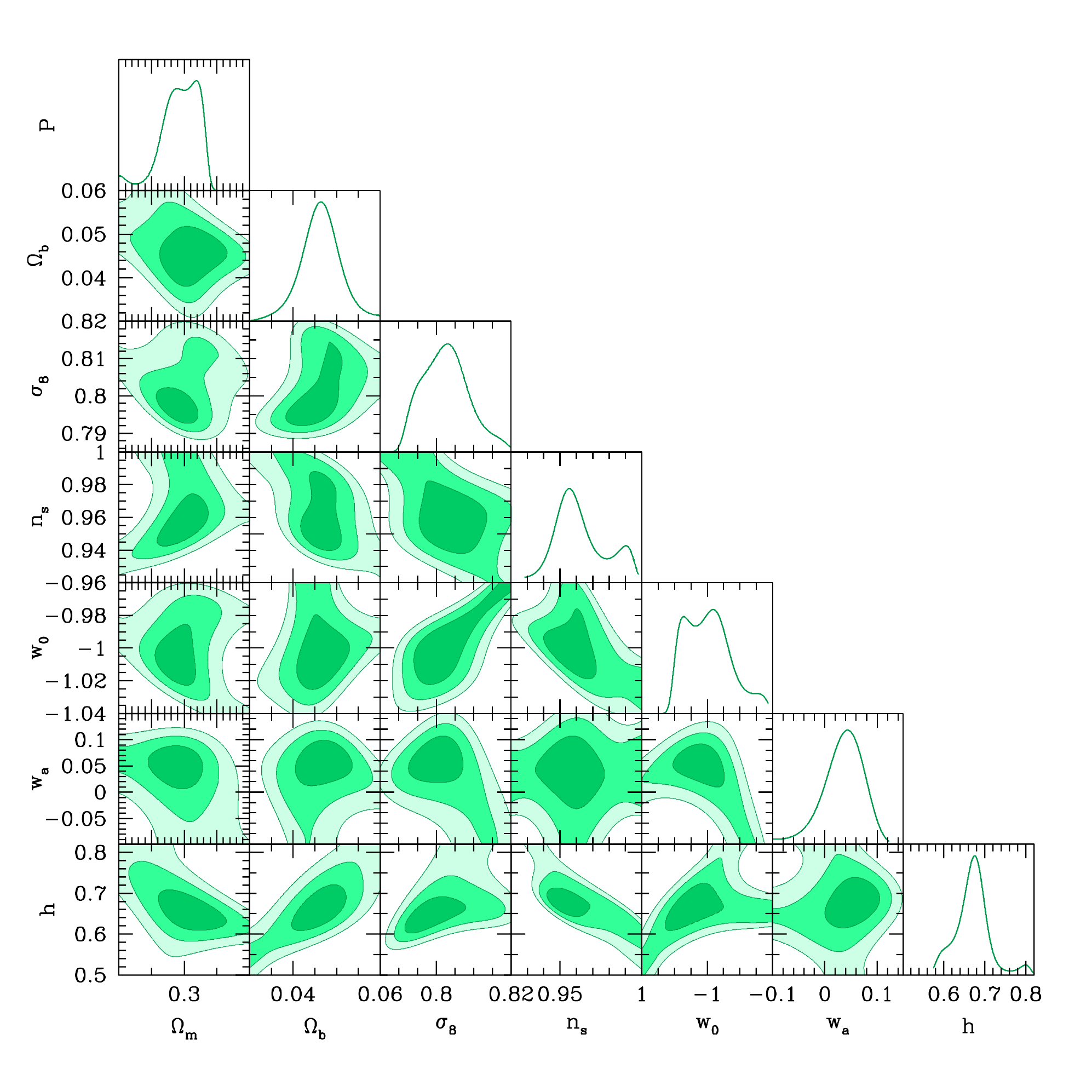}}
}
\caption{The joint constraint results of the seven cosmological parameters from the CSST WL, galaxy clustering, and galaxy-galaxy lensing surveys, assuming $\bar{N}_{\rm add}=10^{-9}$, $f^{z_{\rm s},0}_{\rm eff}=0.5$, and $\bar{N}^g_{\rm sys}=5\times10^4$ $({\rm Mpc}/h)^3$. The $1\sigma$ $(68.3\%)$, $2\sigma$ $(95.5\%)$, and $3\sigma$ $(99.7\%)$ C.L. are shown.}
\label{fig:WL_RSD_all}
\end{figure*}

The constraint results of seven cosmological parameters, i.e. $\Omega_{\rm m}$, $\Omega_{\rm b}$, $\sigma_8$, $n_{\rm s}$, $w_0$, $w_a$, and $h$, from the CSST WL, galaxy clustering, and joint surveys with moderate systematic assumption are shown in Figure~\ref{fig:WL_all}, \ref{fig:RSD_all}, and \ref{fig:WL_RSD_all}.

\end{document}